\RequirePackage{fix-cm} 
\documentclass[a4paper, twoside, reqno, dvips, 12pt]{amsart}
\usepackage{fixltx2e}   

\usepackage{etex}

\usepackage[latin1]{inputenc}
\usepackage[T1]{fontenc}

\usepackage{eucal}
\usepackage{esint}
\usepackage{dsfont}
\usepackage{xspace}
\usepackage{amsgen}
\usepackage{amsthm}
\usepackage{amssymb}
\usepackage{amsmath}
\usepackage{upgreek}
\usepackage{wasysym}
\usepackage{amsfonts}
\usepackage{stmaryrd}
\usepackage{mathtools}

\usepackage{mathrsfs}
\DeclareMathAlphabet{\mathscrbf}{OMS}{mdugm}{b}{n}

\usepackage{a4wide}

\usepackage[dvipsnames, table]{xcolor}
\definecolor{bckg}{RGB}{20.8, 20.8, 20.8}
\definecolor{oneblue}{rgb}{0.0, 0.0, 0.85}
\definecolor{Lightblue}{RGB}{214, 214, 214}
\definecolor{bluepigment}{rgb}{0.2, 0.2, 0.6}
\definecolor{charcoal}{rgb}{0.21, 0.27, 0.31}
\definecolor{denimblue}{rgb}{0.08, 0.38, 0.74}
\definecolor{Lightgray}{rgb}{0.89, 0.89, 0.89}
\definecolor{darkgrey}{rgb}{0.273, 0.281, 0.30}
\definecolor{darkelectricblue}{rgb}{0.33, 0.41, 0.47}

\usepackage{psfrag}
\usepackage{graphicx}
\usepackage{subfigure}
\usepackage{morefloats}
\usepackage{indentfirst}

\usepackage{acronym}
\usepackage{microtype}
\usepackage[labelsep=period,%
            labelfont={bf,sf,color=bluepigment},%
            justification=raggedright]{caption}

\usepackage[perpage, symbol]{footmisc}

\usepackage[usenames, dvipsnames]{pstricks}
\usepackage{epsfig}
\usepackage{pst-grad} 
\usepackage{pst-plot} 

\usepackage[colorlinks,
           urlcolor=oneblue,
           linkcolor=denimblue,
           citecolor=NavyBlue,
           bookmarksopen=false,
           pdfpagemode=UseNone,
           pagebackref]{hyperref}

\usepackage[sort&compress, comma, square, numbers]{natbib}

\usepackage[explicit]{titlesec}

\titleformat{\section}
  {\color{NavyBlue}\Large\sffamily\bfseries}
  {}
  {0em}
  {\colorbox{bckg!5}{\parbox{\dimexpr\linewidth-2\fboxsep\relax}{\centering\thesection. #1}}}
  [\vspace*{0.33em}]

\titleformat{name=\section,numberless}
  {\color{NavyBlue}\Large\sffamily\bfseries}
  {}
  {0.0em}
  {\colorbox{bckg!5}{\parbox{\dimexpr\linewidth-2\fboxsep\relax}{\centering#1}}}
  [\vspace*{0.33em}]

\titleformat{\subsection}
  {\color{NavyBlue}\large\sffamily\bfseries}
  {}
  {0.0em}
  {\colorbox{bckg!5}{\parbox{\dimexpr\linewidth-2\fboxsep\relax}{\centering\thesubsection. #1}}}
  [\vspace*{0.33em}]

\titleformat{name=\subsection,numberless}
  {\color{NavyBlue}\Large\sffamily\bfseries}
  {}
  {0em}
  {\colorbox{bckg!5}{\parbox{\dimexpr\linewidth-2\fboxsep\relax}{\centering#1}}}
  [\vspace*{0.33em}]

\titleformat{\subsubsection}
  {\color{bluepigment}\sffamily\normalsize\bfseries}
  {\thesubsubsection}
  {0.5em}
  {#1}
  [\vspace*{0.33em}]

\titleformat{\paragraph}[runin]
  {\color{bluepigment}\sffamily\small\bfseries}
  {}
  {0em}
  {#1}

\titlespacing{\section}{1.0em}{1.5em plus 2pt minus 2pt}%
{1.0em plus 2pt minus 2pt}[0em]
\titlespacing{\subsection}{1.0em}{1.5em plus 2pt minus 2pt}%
{1.0em}[0em]
\titlespacing{\subsubsection}{1.0em}{1.5em plus 2pt minus 2pt}%
{1.0em plus 2pt minus 2pt}[0em]

\usepackage{titletoc}

\contentsmargin{0.5em}
\newlength{\tocsep} 
\titlecontents{section}[\tocsep]
  {\addvspace{10pt}\bfseries\sffamily}
  {\contentslabel[\thecontentslabel]{\tocsep}}
  {}
  {\ \titlerule*[0.75pc]{.}\ \thecontentspage}
  []
\titlecontents{subsection}[\tocsep]
  {\addvspace{8pt}\sffamily}
  {\contentslabel[\thecontentslabel]{\tocsep}}
  {}
  {\ \titlerule*[0.5pc]{.}\ \thecontentspage}
  []
\titlecontents*{subsubsection}[\tocsep]
  {\addvspace{2pt}\footnotesize\sffamily}
  {}
  {}
  {\ \titlerule*[0.35pc]{.}\ \thecontentspage}
  [\\*]

\makeatletter
\def\@setauthors{%
  \begingroup
  \def\thanks{\protect\thanks@warning}%
  \trivlist
  \centering\footnotesize \@topsep30\p@\relax
  \advance\@topsep by -\baselineskip
  \item\relax
  \author@andify\authors
  \def\\{\protect\linebreak}%
  \textsc{\normalsize\textcolor{darkelectricblue}{\authors}}%
  \ifx\@empty\contribs
  \else
    ,\penalty-3 \space \@setcontribs
    \@closetoccontribs
  \fi
  \endtrivlist
  \endgroup
}
\def\@settitle{\begin{center}%
  \baselineskip14\p@\relax
    \bfseries
    \textsc{\Large\textcolor{charcoal}{\@title}}
  \end{center}%
}
\makeatother

\usepackage{enumitem}
\setlist[description]{%
  topsep=30pt,               
  itemsep=5pt,               
  font={\bfseries\sffamily\color{NavyBlue}}, 
}

\usepackage{fancyhdr}
\usepackage{lastpage}

\newcommand*\Title{\textcolor{bluepigment}{Nonlinear long water waves with surface tension}}
\newcommand*\Authors{\textcolor{bluepigment}{D.~Dutykh, M.~Hoefer \& D.~Mitsotakis}}
\newcommand*{\plogo}{\textcolor{gray}{{\texttt{arXiv.org} / \textsc{hal}}}} 

\numberwithin{equation}{section}

\newtheorem{remark}{Remark}

\newcommand{\ud}{\mathrm{d}}
\newcommand{\ue}{\mathrm{e}}
\newcommand{\Rr}{\mathcal{R}}
\renewcommand{\O}{\mathcal{O}}
\renewcommand{\geq}{\geqslant}
\renewcommand{\leq}{\leqslant}

\renewcommand{\sim}{\thicksim}

\newcommand{\sech}{\mathrm{sech}}

\newcommand{\abs}[1]{\lvert\, #1\, \rvert}

\newcommand{\cf}{\emph{cf.}\xspace}
\newcommand{\ie}{\emph{i.e.}\xspace}
\newcommand{\eg}{\emph{e.g.}\xspace}

\newcommand{\gSerre}{g\textsc{Serre}\xspace}

\newcommand{\dprime}{\prime\prime}

\newcommand{\half}{{\textstyle{1\over2}}}
\newcommand{\third}{{\textstyle{1\over3}}}

\begin{document}

\title[\Title]{Solitary wave solutions and their interactions for fully nonlinear water waves with surface tension in the generalized Serre equations}

\author[D.~Dutykh]{Denys Dutykh}
\address{\textbf{D.~Dutykh:} Univ. Grenoble Alpes, Univ. Savoie Mont Blanc, CNRS, LAMA, 73000 Chamb\'ery, France and LAMA, UMR 5127 CNRS, Universit\'e Savoie Mont Blanc, Campus Scientifique, F-73376 Le Bourget-du-Lac Cedex, France}
\email{Denys.Dutykh@univ-smb.fr}
\urladdr{http://www.denys-dutykh.com/}

\author[M.~Hoefer]{Mark Hoefer}
\address{\textbf{M.~Hoefer:} Department of Applied Mathematics, University of Colorado, Boulder, CO, 80309, USA}
\email{hoefer@colorado.edu}
\urladdr{http://www.colorado.edu/amath/mark-hoefer/}

\author[D.~Mitsotakis]{Dimitrios Mitsotakis$^*$}
\address{\textbf{D.~Mitsotakis:} Victoria University of Wellington, School of Mathematics, Statistics and Operations Research, PO Box 600, Wellington 6140, New Zealand}
\email{dmitsot@gmail.com}
\urladdr{http://dmitsot.googlepages.com/}
\thanks{$^*$ Corresponding author}



\begin{titlepage}
\thispagestyle{empty} 
\noindent
{\Large Denys \textsc{Dutykh}}\\
{\it\textcolor{gray}{CNRS, Universit\'e Savoie Mont Blanc, France}}
\\[0.02\textheight]
{\Large Mark \textsc{Hoefer}}\\
{\it\textcolor{gray}{University of Colorado, USA}}
\\[0.02\textheight]
{\Large Dimitrios \textsc{Mitsotakis}}\\
{\it\textcolor{gray}{Victoria University of Wellington, New Zealand}}
\\[0.16\textheight]

\colorbox{Lightblue}{
  \parbox[t]{1.0\textwidth}{
    \centering\huge\sc
    \vspace*{0.7cm}
    
    \textcolor{bluepigment}{Solitary wave solutions and their interactions for fully nonlinear water waves with surface tension in the generalized Serre equations}

    \vspace*{0.7cm}
  }
}

\vfill 

\raggedleft     
{\large \plogo} 
\end{titlepage}


\newpage
\thispagestyle{empty} 
\par\vspace*{\fill}   
\begin{flushright} 
{\textcolor{denimblue}{\textsc{Last modified:}} \today}
\end{flushright}


\newpage
\thispagestyle{empty} 
\par\vspace*{\fill}   
\begin{flushright} 
{\textcolor{denimblue}{\textsc{Last modified:}} \today}
\end{flushright}


\newpage
\maketitle
\thispagestyle{empty}


\begin{abstract}

Some effects of surface tension on fully-nonlinear, long, surface water waves are studied by numerical means. The differences between various solitary waves and their interactions in subcritical and supercritical surface tension regimes are presented. Analytical expressions for new peaked traveling wave solutions are presented in the dispersionless case of critical surface tension. Numerical experiments are performed using a high-accurate finite element method based on smooth cubic splines and the four-stage, classical, explicit \textsc{Runge--Kutta} method of order four.

\bigskip
\noindent \textbf{\keywordsname:} \textsc{Serre} equations; solitary waves; surface tension; peakons \\


\end{abstract}


\newpage
\tableofcontents
\thispagestyle{empty}


\clearpage
\section{Introduction}
\label{intro}

Waves propagating on the surface of a thin fluid layer can be influenced by the effects of surface tension. Theoretical and experimental studies have shown that interesting phenomena can emerge during the propagation of waves in the presence of surface tension on various fluids including the two extreme cases of water and liquid mercury, \cite{Packham1968, Myers1998, Falcon2002}. For example, in laboratory experiments with thin layers of liquid mercury such that its depth is small compared to the capillary length, elevation and depression solitary waves  have been observed around the critical surface tension regime, \cite{Falcon2002}. For these reasons the mathematical modeling of waves in the presence of surface tension has attracted the interest of scientists that have derived several asymptotic model equations and explored the dynamics of their solutions, \cf \cite{Hunter1988, Amick1989a, Haragus1996, Haragus-Courcelle1998, Daripa2003, Dias2010} and the references therein.

In this paper, we numerically study the effects of surface tension on fully-nonlinear shallow-water waves. Surface water waves are usually described by the full \textsc{Euler} equations of water wave theory, \cite{Whitham1999}. Due to their complexity and the difficulties arising in their theoretical and numerical study, simpler model equations have been derived as approximations to the \textsc{Euler} equations in the shallow water regime. There are three often studied regimes within shallow water waves: (i) the weakly nonlinear -- weakly dispersive, (ii) the weakly nonlinear -- fully dispersive and (iii) the fully nonlinear -- weakly dispersive regime. Model equations such as weakly-nonlinear \textsc{Boussinesq} systems and \textsc{Whitham} type equations modelling capillary-gravity waves in the regimes (i) and (ii) were derived in \cite{Daripa2003, Dinvay2017} (see also \cite{Dash2002, Hur2015}). These models extend the \textsc{Boussinesq} systems derived for surface waves with no surface tension in \cite{BCS} and, although they incorporate surface tension effects, they are limited to small amplitude waves. Because of this approximation, some effects of surface tension cannot be observed due to the absence of higher-order nonlinear terms. For this reason, the study of higher-order models should be considered. Mathematical models appropriate for water waves with surface tension in the regime (iii) were derived in \cite{Dias2010}. These equations extend the \textsc{Serre} equations \cite{Serre1953a,Serre1953}, incorporating surface tension effects, and are referred to as the generalized \textsc{Serre} (\gSerre) equations, \cf \cite{Dias2010}. For the derivation, justification and generalisations of the model equations for surface water waves in both regimes we refer to \cite{Dias2010, Lannes2013}.  In this paper, we focus on the study of solitary wave solutions of the \gSerre equations.

We note that another higher order effect, dispersion, can play a fundamental role in the near critical surface tension regime, as in the fifth order \textsc{Kawahara} equation \cite{Kawahara1972, Hunter1988} that describes weakly nonlinear, unidirectional shallow water waves. While higher order dispersion without surface tension for the fully nonlinear regime (ii) was recently presented in \cite{Matsuno2015}, its generalized, surface tension counterpart was only noted in \cite{Sprenger2016} and has not been studied. Although additional model equations have been derived recently that incorporate more surface tension effects, \cite{Clamond2015a, Clamond2016b}, we will restrict this study to the effects of surface tension on the solitary wave solutions of the \gSerre equations.

\begin{figure}
  \centering
  \includegraphics[width=0.99\textwidth]{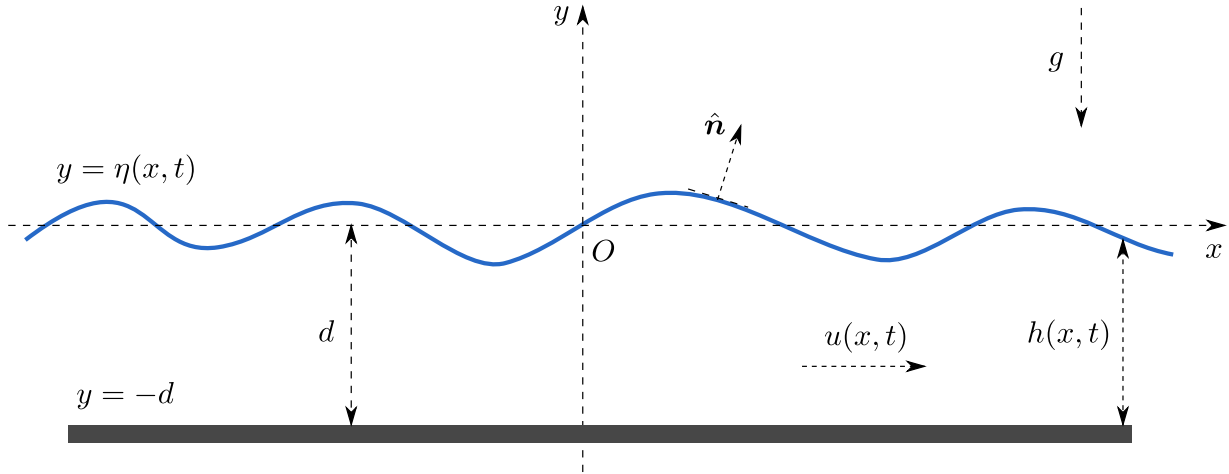}
  \caption{\small\em Sketch of the fluid domain.}
  \label{fig:sketch}
\end{figure}

The \gSerre system of equations is a very accurate mathematical model of shallow water waves that is derived as an approximation to the \textsc{Euler} equations with surface tension and a flat bottom, \cite{Dias2010, Lannes2013}. Here we consider a two-dimensional coordinate system $O\,x\,y$ with the horizontal axis coinciding with the still water level $y\ =\ 0\,$. A layer of a perfect, incompressible fluid (of constant density $\rho\ >\ 0$) is assumed to be bounded between a flat, impermeable bottom at $y\ =\ -\,d$ and its free surface $y\ =\ \eta\,(x,\,t)$ with air. (The variable $\eta$ measures the deviation of the free surface from the still water level.) The \gSerre equations can be obtained by considering the depth-averages horizontal velocity of the fluid and approximating the pressure $p$ jump across the interface using the small slope approximation $\llbracket\,p\,\rrbracket\approx\ -\,\sigma\,\eta_{\,x\,x}\,$. Following the general lines of \cite{Clamond2015c, Mitsotakis2016}, the \gSerre equations for the depth averaged velocity $u\,(x,\,t)$ and the total depth $h\,(x,\,t)\ =\ d\ +\ \eta\,(x,\,t)$ can then be derived using variational methods (first derived in \cite{Dias2010} along with other \gSerre-type equations with different short-wave dispersion properties) in the form:
\begin{align}
 & h_{\,t}\ +\ [\,h\,u\,]_{\,x}\ =\ 0\,, \label{eq:serre1a}\\
 & u_{\,t}\ +\ g\,h_{\,x}\ +\ u\,u_{\,x}\ =\ \frac{1}{3\,h}\;\bigl[\,h^{\,3}\,(u_{\,x\,t}\ +\ u\,u_{\,x\,x}\ -\ u_{\,x}^{\,2})\,\bigr]_{\,x}\ +\ \tau\,h_{\,x\,x\,x}\,.\label{eq:serre2b}
\end{align}
We non-dimensionalise the surface height by the undisturbed water depth $h^{\,\prime}\ =\ h/d\,$, the deviation of the free-surface elevation by a typical wave amplitude $\eta^{\,\prime}\ =\ \eta/a$ the mean velocity by the long wave speed $u^{\,\prime}\ =\ d\,u\,/\,(a\,\sqrt{g\,d})\,$, the horizontal length by a typical wavelegth $x^{\,\prime}\ =\ x/\lambda\,$, and time by the amount of time a long wave takes to traverse a horizontal length $t^{\,\prime}\ =\ \sqrt{g\,d}\,t\,/\,\lambda\,$. Dropping the primes for brevity we obtain the non-dimensional equations
\begin{align}
  & h_{\,t}\ +\ \varepsilon\,[\,h\,u\,]_{\,x}\ =\ 0\,, \label{eq:serre1}\\
  & u_{\,t}\ +\ h_{\,x}\ +\ \varepsilon\,u\,u_{\,x}\ -\ \frac{\delta^{\,2}}{3\,h}\;\left[\,h^{\,3}\,(u_{\,x\,t}\ +\ \varepsilon\,u\,u_{\,x\,x}\ -\ \varepsilon\,u_{\,x}^{\,2})\,\right]_{\,x}\ -\ \delta^{\,2}\,B\,h_{\,x\,x\,x}\ =\ 0\,, \label{eq:serre2}
\end{align}
where $\varepsilon\ =\ a\,/\,d\,$, $\delta\ =\ d\,/\,\lambda\,$. In this setting the total depth from the dimensionless and scaled bottom $z\ =\ -1$ is $h\ =\ 1\ +\ \varepsilon\,\eta$ while both variables $h$ and $u$ are functions of the spatial $x$ and temporal $t$ variables. The coefficient $\tau$ measures the ratio of gravity to capillary forces and is defined as $\tau\ =\ \sigma\,/\,\rho\,$, where $\sigma$ is the surface tension coefficient, $\rho$ the constant density of the fluid. We will utilize the dimensionless \textsc{Bond} number in the form $B\ =\ \tau\,/\,g\,d^{\,2}\ \geq\ 0\,$. Note that this definition of the \textsc{Bond} number, while common in water waves \cite{Dias2010}, is the reciprocal of the \textsc{Bond} number (also called the \textsc{Weber} or \textsc{E\"{o}tv\"{o}s} number) in other areas \cite{Hager2012}. When $B\ =\ 0$ (\ie no surface tension is considered), then the \gSerre equations \eqref{eq:serre1}, \eqref{eq:serre2} reduce to the \textsc{Serre} equations. In the rest of this paper, we focus on the dynamical properties of the model equations where the scaling is not important and so we consider the \gSerre equations in non-dimensional and unscaled form \eqref{eq:serre1} -- \eqref{eq:serre2}, \ie we take $\varepsilon\ =\ \delta^{\,2}\ =\ 1\,$, which is the non-dimensional \gSerre equations \eqref{eq:serre1} -- \eqref{eq:serre2} with $a\ =\ \lambda\ =\ d\,$.

An important, structural property of the \gSerre equations is their linear dispersion relation $\omega^{\,2}\ =\ k^{\,2}\,(1\ +\ B\,k^{\,2})\,/\,(1\ +\ k^{\,2}\,/\,3)\,$. For $B\ =\ 0\,$, the dispersion curvature $\omega^{\,\dprime}\,(k)$ is single-signed for positive wavenumbers $k\,$. The dispersion is convex (or concave). For $B\ >\ 0\,$, the dispersion relation is non-convex, exhibiting an inflection point at $k_{\,\ast}\ =\ \sqrt{3}\,\sqrt{1\ +\ \sqrt{(1\ +\ B)\,/\,B}}\,$. This loss of dispersion convexity can have important physical implications, for example on solitary waves, \cite{Hunter1988, Clamond2015a} and undular bores \cite{Sprenger2016}. This feature makes the \gSerre equations particularly interesting for further study. When $B\ =\ 1/3\,$, the dispersion relation is degenerate $\omega^{\,2}\ =\ k^{\,2}$ and the \gSerre equations are no longer dispersive, designating $B\ =\ 1/3$ as a critical value of surface tension.

Several conservation laws can also be derived for the \gSerre system in a similar way as in the case of the \textsc{Serre} equations. In additions to \eqref{eq:serre1}, here we mention four of these conservation laws: 
\begin{align}\label{eq:cons1}
 & \biggl[\,u\ -\ \frac{(h^{\,3}\,u_{\,x})_{\,x}}{3\,h}\,\biggr]_{\,t}\ +\ \biggl[\,\frac{u^{\,2}}{2}\ +\ g\,h\ -\ \frac{h^{\,2}\,u_{\,x}^{\,2}}{2}\ -\ \frac{u\,(h^{\,3}\,u_{\,x})_{\,x}}{3\,h}\ -\ \tau\,h_{\,x\,x}\,\biggr]_{\,x}\ =\ 0\,, \\
 & \biggl[\,h\,u\ -\ \frac{(h^{\,3}\,u_{\,x})_{\,x}}{3}\,\biggr]_{\,t}\ +\ \biggl[\,h\,u^{\,2}\ +\ \frac{g\,h^{\,2}}{2}\ -\ \frac{2\,h^{\,3}\,u_{\,x}^{\,2}}{3}\nonumber \\
 &\qquad\qquad\qquad\qquad\qquad\qquad\qquad\qquad -\ \frac{h^{\,3}\,u\,u_{\,x\,x}}{3}\ -\ h^{\,2}\,h_{\,x}\,u\,u_{\,x}\ -\ \tau\Rr\,\biggr]_{x}= 0\ ,\label{eq:cons2} \\
 & [\,h\,u\,]_{\,t}\ +\ \bigl[\,h\,u^{\,2}\ +\ \half\;g\,h^{\,2}\ +\ \third\;h^{\,2}\,\gamma\ -\ \tau\,\Rr\,\bigr]_{\,x}\ =\ 0\,, \label{eq:cons3} \\
 & \biggl[\,\frac{h\,u^{\,2}}{2}\ +\ \frac{h^{\,3}\,u_{\,x}^{\,2}}{6}\ +\ \frac{g\,h^{\,2}}{2}\ +\ \frac{\tau}{2}\;h_{\,x}^{\,2}\,\biggr]_{\,t}\ +\ \nonumber\\
 &\qquad\qquad \biggl[\,\Bigl(\frac{u^{\,2}}{2}\ +\ \frac{h^{\,2}\,u_{\,x}^{\,2}}{6}\ +\ g\,h\ +\ \frac{h\,\gamma}{3}\ -\ \tau\,h_{\,x\,x}\Bigr)\,h\,u\ +\ \tau\,h_{\,x}\,(h\,u)_{\,x}\,\biggr]_{\,x}\ =\ 0\,,\label{eq:cons4}
\end{align}
where we introduced two quantities: $\gamma\ =\ h\,\bigl[\,u_{\,x}^{\,2}\ -\
u_{\,x\,t}\ -\ u\,u_{\,x\,x}\,\bigr]\,$, $\Rr\ =\ h\,h_{\,x\,x}\ -\ \half\;h_{\,x}^{\,2}\,$. The quantity $\gamma$ can be considered physically as the vertical acceleration of fluid particles computed at the free surface. It is noted that equation \eqref{eq:cons1} represents the conservation of surface tangential momentum, equations \eqref{eq:cons2} and \eqref{eq:cons3} represent the conservation of horizontal momentum (and are equivalent), and equation \eqref{eq:cons4} represents the conservation of energy.

It is known that the \textsc{Serre} equations admit solitary wave solutions of elevation type (\ie of ${\sech}^2-$type), \cite{Barthelemy2004}. Capillary effects have been shown to be important for the shape of the solitary waves, \cite{Dutykh2016a, Dias2010}. Specifically, the presence of surface tension makes an elevation solitary wave narrower than a solitary wave of the \textsc{Serre} equations with the same speed. Moreover, depending on the value of the \textsc{Bond} number $B\,$, the solitary waves can be either of elevation or of depression type. Depression solitary waves have negative excursion relative to the fluid background. The critical \textsc{Bond} number where the nature of the solitary wave changes, has been found to be $B\ =\ 1/3\,$, \cite{Dias2010}. For $B\ <\ 1/3$ the \gSerre equations admit elevation solitary waves while for $B\ >\ 1/3$ they admit depression solitary waves. It is noted that there are no known analytical formulas for solitary waves of the \gSerre equations when $B\ >\ 0$ and for this reason, numerical computations are required. The change in solitary waves polarity with \textsc{Bond} number near the critical value $B\ =\ 1/3$ has been observed in the full water wave equations too, \cf \eg \cite{Buffoni1996a, Dias1999, Dias1996}. As it was also suggested in \cite{Sprenger2016, Buffoni1996a}, we expect some interesting phenomena around the critical \textsc{Bond} number $B\ =\ 1/3\,$.

The \textsc{Serre} equations ($B\ =\ 0$) were first derived in \cite{Serre1953a, Serre1953} and rederived various times since then, \cf \eg \cite{Su1969}. Although there exist several studies, both theoretical and numerical, for the \textsc{Serre} equations without surface tension \cite{Mirie1982, El2008}, the behaviour of solitary waves under the effects of surface tension remain unknown \cite{Dias2010}. This paper is focused on the properties of solitary wave solutions of the \gSerre equations with surface tension and their interactions and, in general, how surface tension influences their dynamics. Special attention has been given to the critical value of the \textsc{Bond} number $B\ =\ 1/3\,$, where it is shown that the \gSerre equations admit peaked solitary waves of elevation (peakons) and depression (antipeakons) type. The existence of these weakly singular traveling wave solutions along with their stability have been reported in \cite{Mitsotakis2016} while here we provide a detailed description of their construction and their properties. It is noted that the \gSerre equations are the first bi-directional propagation model in water wave theory that admits stable peakons and therefore we can study their head-on collision. The equations \eqref{eq:serre1} -- \eqref{eq:serre2} are solved numerically in a periodic domain using an extension of the standard \textsc{Galerkin}/Finite element method of \cite{Mitsotakis2014} to include the surface tension term. This numerical method has been proven very efficient with favourable convergence properties to smooth and non-smooth solutions \cite{Antonopoulos2017, Mitsotakis2014, Antonopoulos2017a}.

The paper is organized as follows. In Section~\ref{sec:numeth}, we present the numerical methods used in this paper along with their numerical validation. Section~\ref{sec:numex} contains numerical results which analyse the various interactions between elevation and depression solitary waves for multiple values of the \textsc{Bond} number $B\,$. Finally, the main conclusions are discussed in Section~\ref{sec:concl}.


\section{Numerical methods}
\label{sec:numeth}

Consider the \gSerre equations \eqref{eq:serre1} -- \eqref{eq:serre2} in their dimensionless but unscaled form. Lacking analytical expressions for \gSerre solitary waves, we present an efficient numerical method for their numerical generation. Then, the standard \textsc{Galerkin}/finite element method is presented for the numerical integration of the \gSerre equations.


\subsection{Numerical computation of solitary waves}
\label{sec:numint}

It is known that the \textsc{Serre} equations possess solitary wave solutions traveling at constant speed $c_{\,s}$ of the form $h\,(x,\,t)\ =\ 1\ +\ \eta_{\,s}\,(x\ -\ c_{\,s}\,t)\,$, $u\,(x,\,t)\ =\ u_{\,s}\,(x\ -\ c_{\,s}\,t)$ with $\eta_{\,s}\,(\xi)\ =\ A\,{\sech}^{\,2}\,[\,\lambda\,\xi\,]\,$, $u_{\,s}\,(\xi)\ =\ 1\ -\ 1\,/\,(1\ +\ \eta_{\,s}\,(\xi))\,$, $\lambda\ =\ \sqrt{3\,A\,/(4\,(1\ +\ A))}\,$, $c_{\,s}\ =\ \sqrt{1\ +\ A}$ and $\xi\ =\ x\ -\ c_{\,s}\,t\,$.

On the other hand, it was shown in \cite{Dias2010} that the \gSerre equations possess classical solitary wave solutions (symmetric solitary waves that decay monotonically to zero, \cite{DMII}) that satisfy the ordinary differential equation
\begin{equation}\label{eq:desw01}
  (\eta^{\,\prime})^{\,2}\ =\ \eta^{\,2}\;\frac{c_{\,s}^{\,2}\ -\ 1\ -\ \eta}{c^{\,2}_{\,s}\,/\,3\ -\ B\,(1\ +\ \eta)}\,.
\end{equation}
It is noted that the \gSerre equations share the same relation between $u$ and $\eta\,$, \ie $u\ =\ 1\ -\ 1\,/\,(1\ +\ \eta)$ with the \textsc{Serre} equations. Smooth solitary waves also share the same speed-amplitude relation $A\ =\ c_{\,s}^{\,2}\ -\ 1\,$, independent of the \textsc{Bond} number. This is verified by evaluating \eqref{eq:desw01} at the point of the peak amplitude of the solitary wave. In what follows, we describe the numerical method for the computation of the symmetric traveling waves.

Without loss of generality, we search for solitary waves that are positive and symmetric about $x\ =\ 0$ for $0\ \leq\ B\ <\ 1/3\,$. We consider a large enough interval $x\ \in\ [\,-\,L,\,L\,]$ so that the solitary wave has decayed sufficiently close to the background value $\eta\ =\ 0$ within this interval. Under the assumption that $1\ +\ \eta\ \leq\ c_{\,s}^{\,2}\,\min\{1,\,1\,/\,(3\,B)\}$ or $1\ +\ \eta\ \geq\ c_{\,s}^{\,2}\,\max\{1,\,1\,/\,(3\,B)\}$ we can solve the equation \eqref{eq:desw01} for $\eta^{\,\prime}$ and obtain the equation:
\begin{equation}\label{eq:desw1}
  \eta^{\,\prime}\ =\ \pm\,\eta\,\sqrt{\frac{c_{\,s}^{\,2}\ -\ 1\ -\ \eta}{c^{\,2}_{\,s}\,/\,3\ -\ B\,(1\ +\ \eta)}}\,.
\end{equation}
Because of the symmetry of the solution, a nonuniform grid on $[\,0,\,L\,]$ is used, $0\ =\ x_{\,0}\ <\ x_{\,1}\ <\ \cdots\ <\ x_{\,N}\ =\ L\,$, where we assume that the function $\eta$ is decreasing on this grid. Integration of \eqref{eq:desw1} yields
\begin{equation}\label{eq:intdif1}
  -\,\int_{\,\eta\,(0)}^{\,\eta\,(x_{\,i})}\,\frac{1}{\eta}\;\sqrt{\frac{c^{\,2}_{\,s}\,/\,3\ -\ B\,(1\ +\ \eta)}{c_{\,s}^{\,2}\ -\ 1\ -\ \eta}}\;\ud\eta\ =\ \int_{\,0}^{\,x_{\,i}}\ud x\,.
\end{equation}
Making the change of variable $\eta\ =\ \exp\,(z)\,$, then \eqref{eq:intdif1}
can be simplified to the equation
\begin{equation}\label{eq:intdif2}
  \int_{\,\log\,\eta\,(x_{\,i})}^{\,\log\,\eta\,(0)}\,\sqrt{\frac{c^{\,2}_{\,s}\,/\,3\ -\ B\,(1\ +\ \exp\,(z))}{c_{\,s}^{\,2}\ -\ 1\ -\ \exp\,(z)}}\;\ud\,z\ -\ x_{\,i}\ =\ 0\,, 
\end{equation}
which defines the values $\eta_{\,i}\ =\ \eta\,(x_{\,i})$ implicitly, given $\log\,\eta\,(0)\ =\ \log A\,$, $c_{\,s}^{\,2}\ =\ 1\ +\ A\,$, for some $A\,$. \textsc{Gau\textup{\ss}--Legendre} numerical quadrature is used for the approximation of the integral in \eqref{eq:intdif2} while the resulting nonlinear equations are solved with the secant method for values $\eta_{\,i}$ in the interval $(0,\,c_{\,s}^{\,2}\ -\ 1\,]\,$. Usually, the secant method converges in several iterations for a relative error tolerance of $\O\,(10^{\,-\,10})\,$. For the nodes $x_{\,i}\,$, we used the quadrature nodes of the \textsc{Gau\textup{\ss}--Legendre} quadrature rule with $5$ nodes in a uniform grid of the computational domain. It is noted that the discretization of the inner products in the Finite Element method is based on \textsc{Gau\textup{\ss}--Legendre} quadrature and therefore the numerically generated solitary waves can be used directly without using interpolation.

It is noted that in the case $B\ =\ 1/3$ the \gSerre equations are degenerate with no linear dispersion. This implies that the asymptotic approximation of full water waves has broken down and higher order dispersion should be taken into account, \cite{Hunter1988}. In this case, we observe that the equation \eqref{eq:desw1} can be simplified and one can verify that the solution
\begin{equation}\label{eq:peakon}
  \eta_{\,p}\,(\xi)\ =\ A\,\exp(-\,\sqrt{3}\,\abs{\xi})\,,
\end{equation} 
satisfies \eqref{eq:desw1} with $B\ =\ 1/3$ in a trivial way. Therefore, the critical \gSerre equations with $B\ =\ 1/3$ possess peakons of elevation and
depression type since the formula is valid for any $A\,$, \cite{Camassa1993}. Even if solutions of the form \eqref{eq:peakon} satisfy the equation \eqref{eq:desw1} for any value $A\ \in\ \mathds{R}$ it has been found numerically that only the peakons that satisfy the usual speed-amplitude relation $c_{\,s}\ =\ \sqrt{1\ +\ A}$ are stable, \cite{Mitsotakis2016}. Such stable solutions are the limits of smooth solitary waves for $B\ \not=\ 1/3$ as $B\ \rightarrow\ 1/3\,$. This can be seen by making the dependence of the traveling-wave solutions on the parameter $B$ explicit and assuming solitary waves of the form $\eta\ =\ \eta\,(\xi;\,B)$ satisfying \eqref{eq:desw1} with $c_{\,s}\ =\ \sqrt{1\ +\ A}\,$. Then, from \eqref{eq:desw1} after integration  we obtain:
\begin{equation}\label{eq:newsw}
  \eta\,(\xi;\,B)\ =\ A\,\exp\left(\,\pm\int\sqrt{\frac{c_{\,s}^{\,2}\ -\ 1\ -\ \eta}{c_{\,s}^{\,2}\,/\,3\ -\ B\,(1\ +\ \eta)}}\;\ud\,\xi\right)\,,
\end{equation}
which is simplified to the peakon solution 
\begin{equation*}
  \lim_{B\,\to\,1/3}\,\eta\,(\xi;\,B)\ =\ \eta_{\,p}\,(\xi)\ =\ A\,\exp(\,-\,\sqrt{3}\,\abs{\xi})\,,
\end{equation*}
after taking the limit $B\ \rightarrow\ 1/3\,$. The existence of these weakly singular solutions has been demonstrated also using phase-plane analysis of the relevant ordinary differential equation in \cite{Mitsotakis2016}.

\begin{remark}
Formula \eqref{eq:newsw} can provide an asymptotic estimate for the decay rate of the solitary waves of the \gSerre equations. Since solitary waves are assumed to decay at infinity, \ie $\eta\ \ll\ 1$ as $\abs{\xi}\ \to\ \infty\,$, we have from \eqref{eq:newsw} that 
\begin{equation}\label{eq:asym}
  \eta\,(\xi;\,B)\ \sim\ A\,\exp{\left(\,-\sqrt{\frac{c_{\,s}^{\,2}\ -\ 1}{c_{\,s}^{\,2}\,/\,3\ -\ B}}\;\abs{\xi}\right)}\,, \quad \abs{\xi}\ \to\ \infty\,.
\end{equation}
This exponential decay of the solitary waves is also illustrated in Figure~\ref{fig:sws1}.
\end{remark}

\begin{remark}
The asymptotic relation \eqref{eq:asym} can also been obtained from \eqref{eq:desw01} written in the form $(\eta^{\,\prime})^{\,2}\ =\ F\,(\eta)$ and using the second order \textsc{Taylor} expansion of $F\,(\eta)\ \approx\ F^{\,\dprime}\,(0)\,\eta^{\,2}\,/\,2\ +\ \O\,(\eta^{\,3})$ where $F^{\,\dprime}\,(0)\ =\ 2\,(c_{\,s}^{\,2}\ -\ 1)\,/\,(c_{\,s}^{\,2}\,/\,3\ -\ B)\,$.
\end{remark}


\subsection{Numerical time integration of the gSerre equations}

For the numerical approximation of the initial value problem of the \gSerre equations subject to periodic boundary conditions, we implement a standard \textsc{Galerkin} / finite element method for the spatial discretization and the fourth-order, four-stage classical \textsc{Runge--Kutta} method for the discretization in time (\cf \cite{Mitsotakis2014} for the \textsc{Serre} equations lacking surface tension). Consider the system \eqref{eq:serre1} -- \eqref{eq:serre2} posed in a finite interval $[\,a,\,b\,]$ and for time $t\ \in\ [\,0,\,T\,]$ with some $T\ >\ 0\,$, and with periodic boundary conditions, \ie $h^{\,(k)}\,(a,\,t)\ =\ h^{\,(k)}\,(b,\,t)$ and $u^{\,(k)}\,(a,\,t)\ =\ u^{\,(k)}\,(b,\,t)$ for $k\ =\ 0,\,1,\,2,\,\ldots$. We consider a uniform subdivision of the interval $[\,a,\,b\,]$ consisting of the nodes $x_{\,i}\ =\ a\ +\ i\,\Delta\,x\,$, where $i\ =\ 0,\,1,\,\cdots,\,N\,$, and $N\ \in\ \mathds{N}\,$, such that the grid size is defined as $\Delta\,x\ =\ (b\ -\ a)/N\,$. We shall consider numerical solutions of the gSerre equations in the space of cubic, periodic splines
\begin{equation*}
  S\ =\ \{\phi\ \in\ \left.C^{\,2}_{\,\mathrm{per}}\,[\,a,\,b\,]\ \right\vert\ \phi\bigr\vert_{\,[x_{\,i},\,x_{\,i\,+\,1}\,]}\ \in\ \mathds{P}^{\,3}\,, \quad 0\ \leq\ i\ \leq\ N\ -\ 1\}\,,
\end{equation*}
where
\begin{equation*}
  C_{\,\mathrm{per}}^{\,2}\ =\ \{\left.f\ \in\ C^{\,2}\,[\,a,\,b\,]\right\vert\ f^{\,(k)}\,(a)\ =\ f^{\,(k)}\,(b), \quad 0\ \leq\ k\ \leq\ 2\}\,,
\end{equation*}
and $\mathds{P}^{\,3}$ is the space of cubic polynomials.

The numerical solution will be denoted by $\tilde{h}$ and $\tilde{u}\,$. To state the spatial \textsc{Galerkin} semi-discretization, we first multiply equations \eqref{eq:serre1} -- \eqref{eq:serre2} by $\phi\ \in\ S\,$. Integration by parts leads to the weak formulation:
\begin{align}
  & (\tilde{h}_{\,t},\,\phi)\ +\ ((\tilde{h}\,\tilde{u})_{\,x},\,\phi)\ =\ 0\,, \label{eq:weak1}\\
  & \mathcal{B}\,(\tilde{u}_{\,t},\,\phi;\,\tilde{h})\ +\ (\tilde{h}\,(\tilde{h}_{\,x}\ +\ \tilde{u}\,\tilde{u}_{\,x}),\,\phi)\ +\ \frac{1}{3}\;\left(\tilde{h}^{\,3}\,(\tilde{u}\,\tilde{u}_{\,x\,x}\ -\ (\tilde{u}_{\,x})^{\,2}),\,\phi_{\,x}\right)\ +\nonumber \\ 
  & \qquad\qquad B\,\left[\,(\tilde{h}\,\tilde{h}_{\,x\,x},\,\phi_{\,x})\ +\ (\tilde{h}_{\,x}\,\tilde{h}_{\,x\,x}\,,\,\phi)\,\right]\ =\ 0\,, \label{eq:weak2}
\end{align}
where $(f,\,g)\ =\ \int_{\,a}^{\,b}\;f\,g\;\ud x\,$, the bilinear form $\mathcal{B}$ is defined for a fixed $\tilde{h}$ (and $\phi,\,\psi\ \in\ S$) as
\begin{equation}\label{eq:B}
  \mathcal{B}\,(\psi,\,\phi;\,\tilde{h})\ =\ (\tilde{h}\,\psi,\,\phi)\ +\ \frac{1}{3}\;(\tilde{h}^{\,3}\,\psi_{\,x},\,\phi_{\,x})\,,
\end{equation}
and the initial conditions are
\begin{equation}\label{eq:weak-init} 
  \tilde{h}\,(x,\,0)\ =\ P\,h_{\,0}\,(x)\,, \qquad 
  \tilde{u}\,(x,\,0)\ =\ P\,u_{\,0}\,(x)\,,
\end{equation}
where $P$ is the $L^{\,2}$ projection onto $S$ defined by $(P\,v,\,\phi)\ =\ (v,\,\phi)\,$, for all $\phi\ \in\ S\,$.

Using the standard basis functions with B$-$splines for the space $S\,$, the equations \eqref{eq:weak1} -- \eqref{eq:weak2} form a system of ordinary differential equations. It has been shown that the classical, explicit, four-stage, fourth-order \textsc{Runge--Kutta} method performs very well for the surface tensionless \textsc{Serre} equations, \cite{Mitsotakis2014}. Denoting by $\Delta\,t$ the uniform time-step, we consider the temporal grid $t^{\,n}\ =\ n\,\Delta\,t\,$, for $n\ =\ 0,\,1,\,\cdots,\,K\,$, with $\Delta\,t\ =\ T/K\,$. For more information about the formulation and properties of this fully-discrete scheme, we refer the reader to \cite{Mitsotakis2014, Antonopoulos2017}. We only note here that the convergence rates for the present system are analogous to the convergence rates observed for the \textsc{Serre} system while in the case of peakons the convergence properties are similar to the case of the \textsc{Camassa--Holm} equation, \cite{Antonopoulos2017a}. Specifically, the convergence rate of the standard Galerkin method depends on the smoothness of the solutions and therefore the convergence rate for non-smooth peakon solutions can be reduced to order $1$ while for smooth solutions it is of order $4\,$. This is reflected in the conservation properties of the method and so it requires fine grids when dealing with non-smooth solutions. For the numerical computation of the integrals appearing in the numerical method, we use the \textsc{Gau\textup{\ss}--Legendre} quadrature rule with five nodes in each spatial mesh interval. The nodes of the quadrature rule form a nonuniform grid which we used in the numerical method presented in Section~\ref{sec:numint}.


\subsection{Method validation}

\begin{figure}
  \centering
  \bigskip
  \includegraphics[width=0.99\columnwidth]{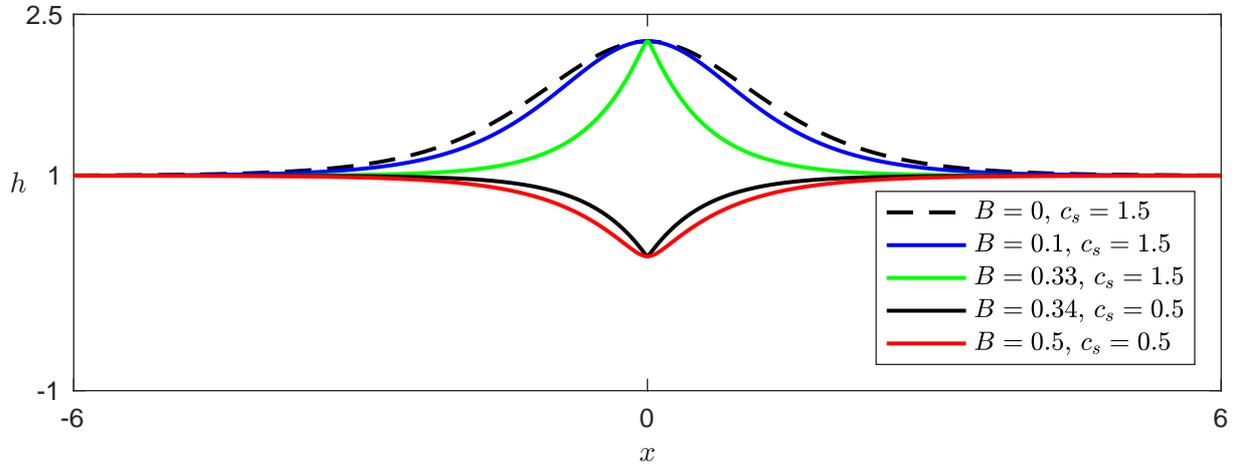}
  \caption{Numerically computed solitary waves for the \gSerre equations with various values of \textsc{Bond} number $B\,$.}
  \label{fig:sws1}
\end{figure}

In order to validate the presented numerical methods for solitary wave computation and time integration, we generate and numerically evolve solitary wave solutions for various values of the \textsc{Bond} number $B\,$. As it was noted in \cite{Dias2010}, solitary waves for $B\ <\ 1/3$ are of elevation type while for $B\ >\ 1/3$ are of depression type, \cf Figure~\ref{fig:sws1}. Solitary waves are computed on the domain $[\,-40,\,40\,]\,$. Figure~\ref{fig:sws1} presents a magnification of the numerically generated solitary waves for various values of the \textsc{Bond} number $B$ along with an exact solitary wave of the \textsc{Serre} system with $c_{\,s}\ =\ 1.5\,$.

The numerical method for the solution of \eqref{eq:desw1} typically converges quickly and the convergence appears to depend slightly on the amplitude of the solitary waves and the value of the \textsc{Bond} number. As an indication of the speed of the numerical method described in Section~\ref{sec:numint}, we present here the elapsed time for the computations of the solitary waves on the \textsc{Gau\textup{\ss}--Legendre} quadrature nodes with $\Delta\,x\ =\ 0.1$ ($N\ =\ 800$) in $[\,-40,\,40\,]\,$. For $B\ =\ 0.1$ and $c_{\,s}\ =\ 1.5\,$, the CPU time is $10.4265$ seconds, while, for $B\ =\ 0.33$ and $c_{\,s}\ =\ 1.5$ is $11.1082$ seconds, very similar to the case $B\ =\ 0.35$ and $c_{\,s}\ =\ 0.5\,$, which is $11.5507$ seconds. As the \textsc{Bond} number approaches the value $B\ =\ 0$ the elapsed time is again close $11$ seconds. For example, for $B\ =\ 0.1$ and $c_{\,s}\ =\ 3$ the elapsed time is $10.9633\,$. In general, we observe that the convergence time of the solitary wave computation is comparable as $B$ and $A$ are varied.

In order to study the accuracy of the approximation of the solitary waves, we use the numerically generated solitary waves for $B\ =\ 0.1$ and $B\ =\ 0.5$ as initial conditions to the fully discrete numerical time integration scheme and study several error indicators relevant to the propagation of traveling waves. Specifically, we monitor the amplitude, speed, shape and phase errors for solitary wave propagation up to time $T\ =\ 100\,$. It is noted that for elevation solitary waves, we use the discretization parameters $\Delta\,x\ =\ 0.1$ and $\Delta\,t\ =\ 0.05\,$, while for depression solitary waves we use smaller mesh lengths $\Delta\,x\ =\ 0.01$ and $\Delta\,t\ =\ 0.005$ in order to produce stable and accurate computations. For the specific numerical method it has been shown that mild restrictions on $\Delta\,t\ \leq\ C\,\Delta\,x$ are adequate.

\begin{figure}
  \centering
  \bigskip
  \includegraphics[width=0.79\columnwidth]{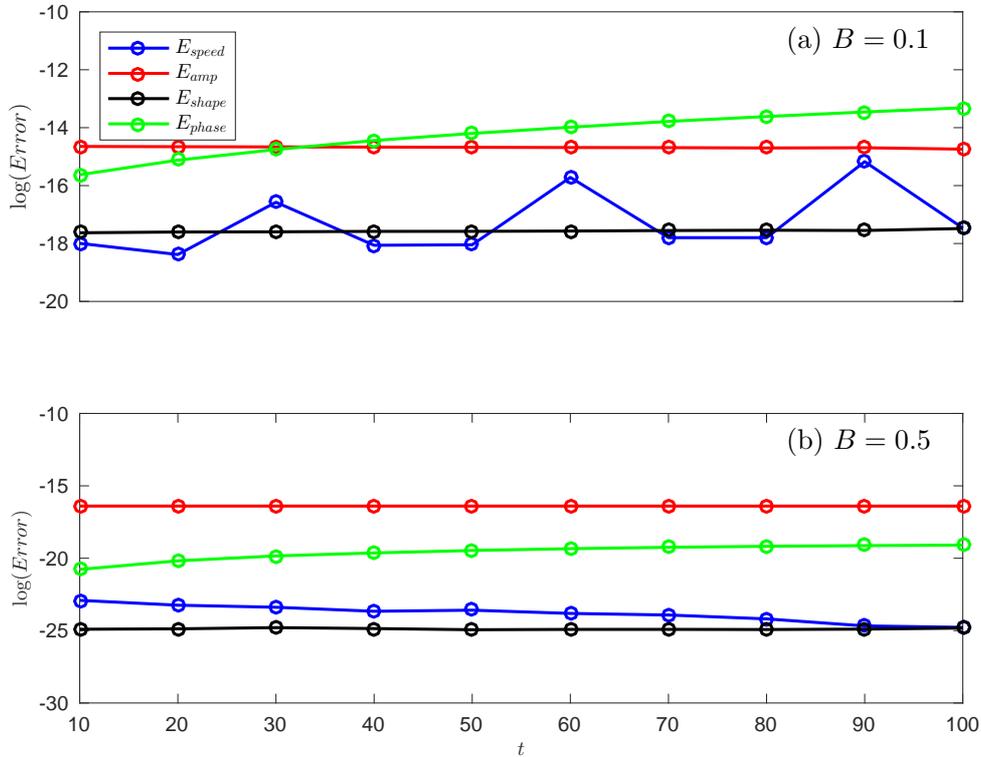}
  \caption{\small\em Error indicators for the propagation of two solitary waves.}
  \label{fig:errs}
\end{figure}

We define the normalized amplitude error as
\begin{equation}\label{eq:AE}
  E_{\,\mathrm{amp}}\ =\ \frac{\abs{H\,(x^{\,\ast}\,(t),\,t)\ -\ H_{\,0}}}{\abs{H_{\,0}}}\,,
\end{equation}
where $x^{\,\ast}\,(t)$ is the curve along which the computed approximate solution $H\,(x,\,t)$ achieves its maximum and $H_{\,0}\ \equiv\ H\,(0)$ is the initial peak amplitude of the numerically generated solitary wave. We observe that $E_{\,\mathrm{amp}}$ remains very small and practically constant during the propagation of the solitary waves, \cf~Figure~\ref{fig:errs}.

Additionally, we approximate the solitary wave speed $c_{\,s}$ by $\tilde{c}_{\,s}$ as
\begin{equation}\label{eq:spped}
  \tilde{c}_{\,s}\ =\ \frac{x^{\,\ast}\,(t)\ -\ x^{\,\ast}\,(t\ -\ \tau)}{\tau}\,,
\end{equation}
where $\tau$ is a constant. The corresponding speed error is defined as
\begin{equation*}
  E_{\,\mathrm{speed}}\ =\ \frac{\abs{\tilde{c}_{\,s}\ -\ c_{\,s}}}{\abs{c_{\,s}}}\,.
\end{equation*}
The results for $\tau\ =\ 10$ in Figure~\ref{fig:errs} show that the error between the numerical values $\tilde{c}_s$ and the exact speed $c_{\,s}$ remained practically constant.

Two other error norms that are pertinent to solitary waves are the \emph{shape and phase errors}. We define the normalized shape error as the distance in $L^{\,2}$ between the computed solution at time $t\ =\ t^{\,n}$ and the family of translated, exact solitary waves with the same amplitude/speed, \ie,
\begin{equation}\label{eq:shape}
  E_{\,\mathrm{shape}}\ =\ \min_\tau \zeta\,(\tau)\,, \quad 
  \zeta\,(\tau)\ =\ \frac{\abs{H\,(x,\,t^{\,n})\ -\ h\,(x,\,\tau)}}{\abs{h\,(x,\,0)}}\,.
\end{equation}
The minimum in \eqref{eq:shape} is attained at some critical $\tau\ =\ \tau^{\,\ast}\,(t^{\,n})\,$. This, in turn, is used to define the (signed) phase error as
\begin{equation}\label{eq:phase}
  E_{\,\mathrm{phase}}\ =\ \tau^{\,\ast}\ -\ t^{\,n}\,.
\end{equation}
In order to find $\tau^{\,\ast}\,$, we solve the equation $\zeta^{\,\prime}\,(\tau)\ =\ 0$ using \textsc{Newton}'s method. The initial guess for \textsc{Newton}'s method is chosen as $\tau^{\,0}\ =\ t^{\,n}\ -\ \Delta t\,$. Having computed $\tau^{\,\ast}\,$, the shape error \eqref{eq:shape} is then
\begin{equation*}
  E_{\,\mathrm{shape}}\ =\ \zeta(\tau^{\,\ast})\,.
\end{equation*}
These error norms are closely related to the \emph{orbit} of the solitary wave (family of translated solitary waves) and measure properties of the solitary waves, which are often not well conserved using dissipative numerical methods.

The computed numerical errors are presented on a logarithmic scale in Figure~\ref{fig:errs}. It can be observed that the errors in the propagation of both solitary waves are very small. Especially when we use grids with small $\Delta x\,$, the numerical speed of propagation of the solitary wave is almost equal to the exact speed and similarly the rest of the error indicators remain very small. This study shows that both numerical methods are very accurate and they conserve the properties of a propagating solitary wave very well. We note that the errors are analogous for solitary waves with different propagation speeds since they primarily depend on the discretization $\Delta x$ and $\Delta t\,$. We also mention that the fully-discrete scheme appears to be stable with no restrictive bounds on the ratio $\Delta t/\Delta x$ except when the solution is not very smooth. Mild restrictions, empirically found to be on the order $\Delta t/\Delta x\ \lessapprox\ 10^{\,-\,2}\,$, could be necessary for the numerical stability even of the non-smooth solutions.

As the \textsc{Bond} number $B$ approaches the value $1/3\,$, solitary wave solutions become more cusp-shaped, approaching the peakon solution~\eqref{eq:peakon}, \cf Figure~\ref{fig:sws1}. For example, the solitary waves for $B\ =\ 0.33$ are very close to peaked solitary waves, \cite{Camassa1993, Liao2014}. This phenomenon has already been explained in Section~\ref{sec:numint} and we verify it here numerically. As the solitary waves lose smoothness, then the mesh length $\Delta x$ must be reduced in order to maintain high resolution. We discuss further the transcritical case where $B\ \approx\ 1/3$ in Section~\ref{sec:numex}.

In order to study the conservation properties of the numerical scheme, we consider the following conserved quantities:
\begin{align*}
  E_{\,1}\,(t)\ &=\ \int_{\,a}^{\,b} \left(u\ -\ h\,h_{\,x}\,u_{\,x}\ -\ \frac{h^{\,2}\,u_{\,x\,x}}{3}\right)\ud x\,, \\
  E_{\,2}\,(t)\ &=\ \int_{\,a}^{\,b}\left(h\,u\ -\ h^{\,2}\,h_{\,x}\,u_{\,x}\ -\ \frac{h^{\,3}\,u_{\,x\,x}}{3}\right)\ud x\,,\\
  E_{\,3}\,(t)\ &=\ \int_{\,a}^{\,b} h\,u \ud x\,, \\
  E_{\,4}\,(t)\ &=\ \int_{\,a}^{\,b} \left(h\,u^{\,2}\ +\ \frac{h^{\,3}\,u_{\,x}^{\,2}}{3}\ +\ h^{\,2}\ +\ B\,h_{\,x}^{\,2}\right)\ud x\,.
\end{align*}
These quantities occur after integration of the conservation laws \eqref{eq:cons1} -- \eqref{eq:cons4} over a periodic interval $[\,a,\,b\,]$ while taking into account the periodic boundary conditions. We also consider the evolution of initial conditions of the form $\eta\,(x,\,0)\ =\ C\,\ue^{\,-\,0.1\,x^{\,2}}$ and $u\,(x,\,0)\ =\ 0$ in the interval $[\,-100,\,100\,]$ up to $T\ =\ 100$ for various values of the parameter $B\,$. In general the quantities $E_{\,i}\,(t)\,$, $i\ =\ 1,\,2,\,3$ are conserved in all of our simulations with at least $11$ digits correct for all values of $B$ and $C$ we used. For example, when we take $\Delta x\ =\ 0.1$ and $\Delta t\ =\ 0.01\,$, the quantities $\abs{E_{\,i}\,(t)\ -\ E_{\,i}\,(0)}$ for $i\ =\ 1,\,2,\,3$ are always of $\O\,(10^{\,-\,10})\,$. The quantity $E_{\,4}$ is well conserved for small and large values of $B$ but is poorly conserved for values of $B$ close to the critical value $1/3$ due to the singularity in the first derivative of the solution. For example, for $B\ =\ 0.1$ and $C\ =\ 1$ the conserved value of $E_{\,4}$ is $214.9750606245\,$. For $B\ =\ 0.5$ and $C\ =\ -0.8$ the conserved value of $E_{\,4}$ is $195.4778205\,$. For $B\ =\ 1/3$ although the quantities $E_{\,i}$ for $i\ =\ 1,\,2,\,3$ are conserved again; the value $E_{\,4}$ is conserved only with $5$ digits and is $215.05\,$. Taking smaller values for $\Delta x$ and $\Delta t\,$, the conserved quantities can be improved. This behaviour is expected since the convergence of the \textsc{Galerkin} method depends on the regularity properties of the solution. The results indicate that the \textsc{Galerkin} method preserves the quantities $E_{\,i}$ for $i\ =\ 1,\,2,\,3$ while the presence of peaked solitary waves affect the conservation of the quantity $E_{\,4}\,$.


\section{Numerical experiments}
\label{sec:numex}

In this section, we study the effects of surface tension on various solitary wave interactions. Specifically, we study the head-on and overtaking collisions of elevation and depression solitary waves. We also study the generation and interaction of solitary waves when the \textsc{Bond} number $B\ \approx\ 1/3\,$. It is noted that in what follows, we will report the deviation of the free surface $\eta$ rather than the total depth $h\,$. Thus, we consider initial conditions that are less than $10^{\,-\,10}$ on the boundaries of the computational domain.


\subsection{Head-on collisions}
\label{sec:hoc}

\begin{figure}
  \centering
  \bigskip
  \includegraphics[width=0.9\columnwidth]{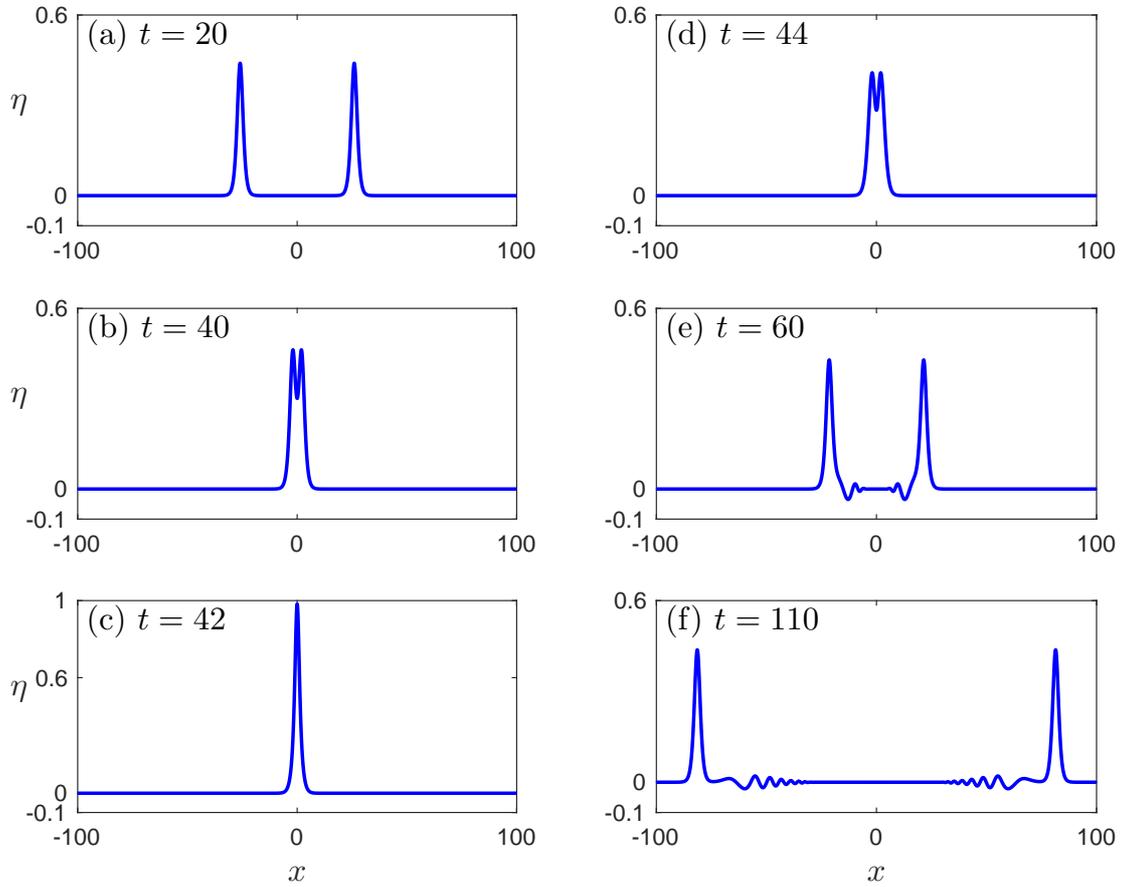}
  \caption{\small\em Symmetric head-on collision of two elevation solitary waves for the \gSerre equations with $B\ =\ 0.1\,$.}
  \label{fig:col1}
\end{figure}

We first study the symmetric head-on collision for the \gSerre equations with $B\ =\ 0.1$ for two identical elevation solitary waves that propagate in opposite directions. In these numerical experiments, we consider the interval $[\,-200,\,200\,]$ and take $\Delta x\ =\ 0.1\,$, $\Delta t\ =\ 0.01\,$. Here we present the solitary waves with $c_{\,s}\ =\ 1.2$ and amplitude $A\ =\ 0.4472\,$. The solitary waves are initially translated so that their maximum values are achieved at $x\ =\ -\,50$ and $x\ =\ 50\,$, respectively, exhibiting essentially no overlap in their exponentially small tails. The interaction begins at approximately $t\ =\ 40$ and the peak of the interaction occurs at about $t\ =\ 42\,$. The interaction is presented in Figure~\ref{fig:col1}. We observe that after the collision, the solitary waves propagate in different directions followed by small amplitude dispersive tails.

\begin{figure}
  \centering
  \bigskip
  \includegraphics[width=0.9\columnwidth]{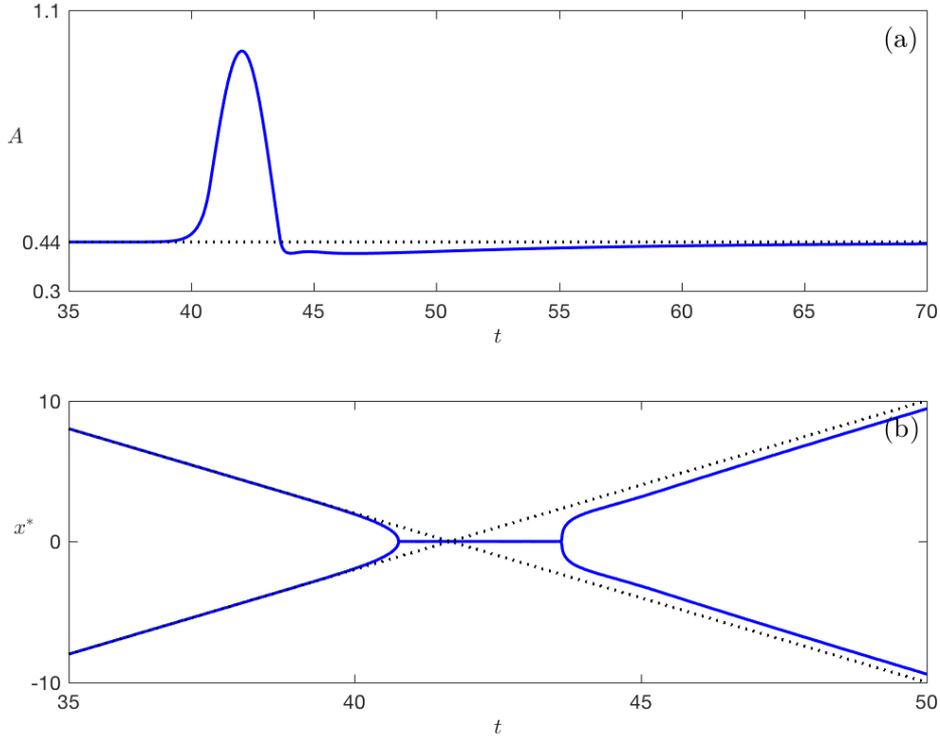}
  \caption{\small\em (a) Peak amplitude of the solution as a function of time, (b) Phase diagram of the location of the solitary waves during the interaction of Figure~\ref{fig:col1}.}
  \label{fig:col2}
\end{figure}

In addition to the generation of dispersive tails, the inelastic interaction causes a phase change in the propagation of the solitary waves. During the collision there is a temporal interval in which the solution has only one peak while the maximum value of the solution during the interaction recorded is $0.9838\,$, which is greater than the sum of the amplitudes, $0.88\,$, of the solitary waves. The solitary waves after the interaction are separated and stabilized to different amplitudes $A\ \approx\ 0.4377$ compared with the initial amplitudes $A\ =\ 0.4472\,$.

Figure~\ref{fig:col2} shows the amplitude (\textit{a}) and the location of the maximum values of the solution (\textit{b}) recorded during the interaction. We observe that, due to nonlinear interaction, the amplitude fluctuates before it is stabilized to its new value. The dotted lines in these diagrams represent the predicted solitary wave maximum for no interaction.

\begin{figure}
  \centering
  \bigskip
  \includegraphics[width=0.9\columnwidth]{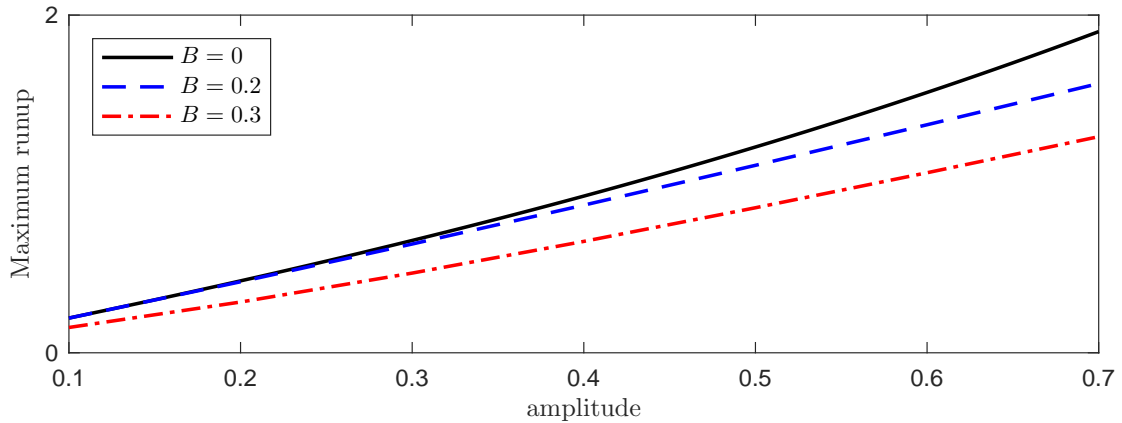}
  \caption{\small\em Maximum runup of solitary waves for $B\ =\ 0\,$, $0.2$ and $0.3\,$.}
  \label{fig:rup}
\end{figure}

We perform several symmetric head-on collisions and record the maximum value at $x\ =\ 0$ during the interaction. This value is also known as the maximum runup since, by reflection symmetry $h\,(x,\,t)\ =\ h\,(-x,\,t)\,$, $u\,(x,\,t)\ =\ -\,u\,(x,\,t)$ of the equations and the initial data, it coincides with the maximum runup of a solitary wave on a vertical wall located at $x\ =\ 0$ subject to appropriate boundary conditions \cite{Su1980, Mirie1982}. Figure~\ref{fig:rup} shows the values recorded for $B\ =\ 0.2$ and $0.3$ compared with the asymptotic solution of \cite{Mirie1982} for $B\ =\ 0\,$. We observe that surface tension decreases the maximum runup value. This effect is stronger for larger amplitude solitary waves for the same values of $B\,$. Moreover, the maximum runup value is decreasing as $B$ is increasing. Therefore, for larger values of $B\,$, we observe smaller maximum runup values.

\begin{figure}
  \centering
  \bigskip
  \includegraphics[width=0.99\columnwidth]{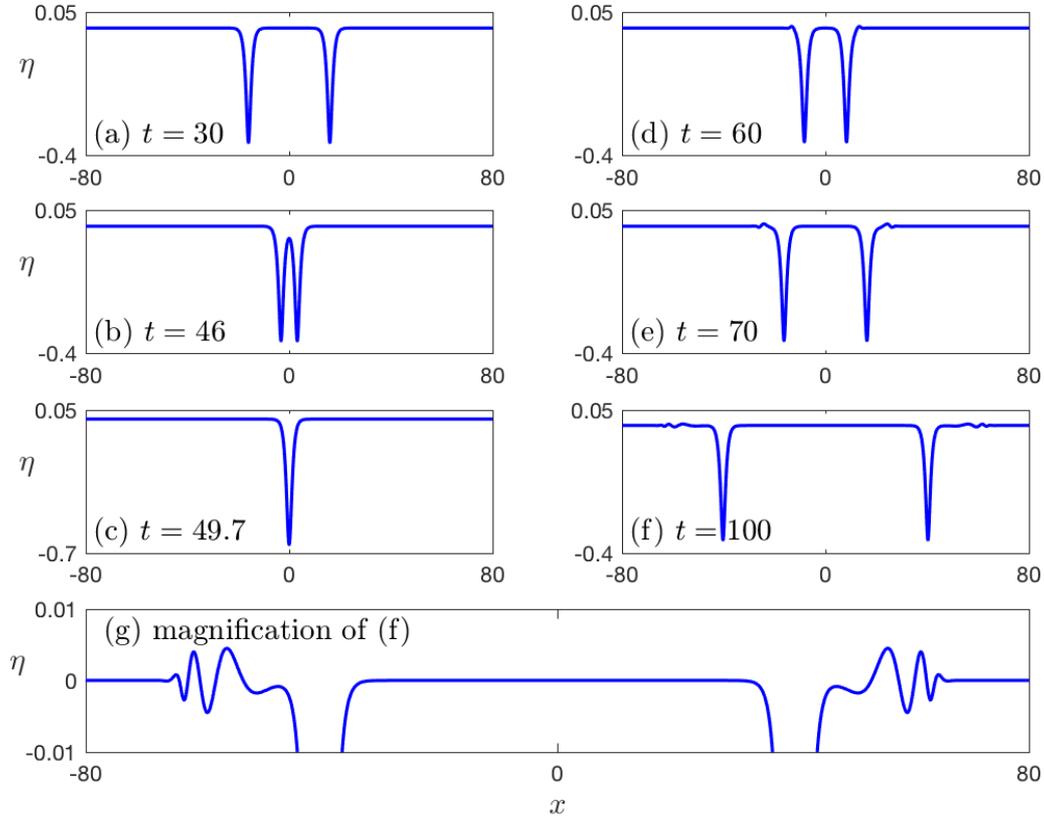}
  \caption{\small\em Symmetric head-on collision of two depression solitary waves for the \gSerre equations with $B\ =\ 0.5\,$.}
  \label{fig:col3}
\end{figure}

We also study the unsymmetric collision of two solitary waves with amplitudes $A\ =\ 0.96$ and $0.44\,$, respectively (equivalently, speeds $1.4$ and $1.2\,$, respectively). The interaction is very similar to the unsymmetric interaction of the \textsc{Serre} equations and is not presented here. We only mention that the solitary waves after the interaction have amplitudes $A\ \approx\ 0.9515$ and $0.4325$ respectively. In conclusion, the head-on collision of elevation solitary waves with small surface tension is qualitatively the same as the collision of solitary waves with no surface tension. It is noted that a jet formation can be observed during the head-on collision of large amplitude solitary waves, \cite{Chen2015a, Mitsotakis2014}. This has been observed in numerical experiments with the \textsc{Euler} equations while with the \textsc{Serre} equations can be observed a weak jet formation during the head-on collision of large amplitude solitary waves \cite{Mitsotakis2014} but the exact jet formation is not possible for the \textsc{Serre} equations since the water height must be a single-valued function.

We continue with the head-on collision of depression solitary waves for \textsc{Bond} number greater than $1/3\,$. The interaction of depression solitary waves has never been studied before and there are no previous results to compare with. We study the symmetric head-on collision of two solitary waves with speed $c_{\,s}\ =\ 0.8$ and amplitude $A\ =\ -\,0.36$ (where we keep the minus sign to emphasise that the solution is negative) for \textsc{Bond} number $B\ =\ 0.5\,$. Initially, the solitary waves are translated to $x\ =\ -\,40$ and $40$ respectively and we solve the \gSerre equations in the interval $[\,-100,\,100\,]$ with $\Delta x\ =\ 0.01$ and $\Delta t\ =\ 0.001\,$. Figure~\ref{fig:col3} shows the inelastic interaction between two depression solitary waves. It is interesting that the dispersive tails generated after the interaction propagate faster than the solitary waves and therefore lead in the propagation while there are no deviations of the free surface between the two new solitary pulses. This is due to the linear dispersion curvature sign change as $B$ crosses $1/3\,$.

\begin{figure}
  \centering
  \bigskip
  \includegraphics[width=0.9\columnwidth]{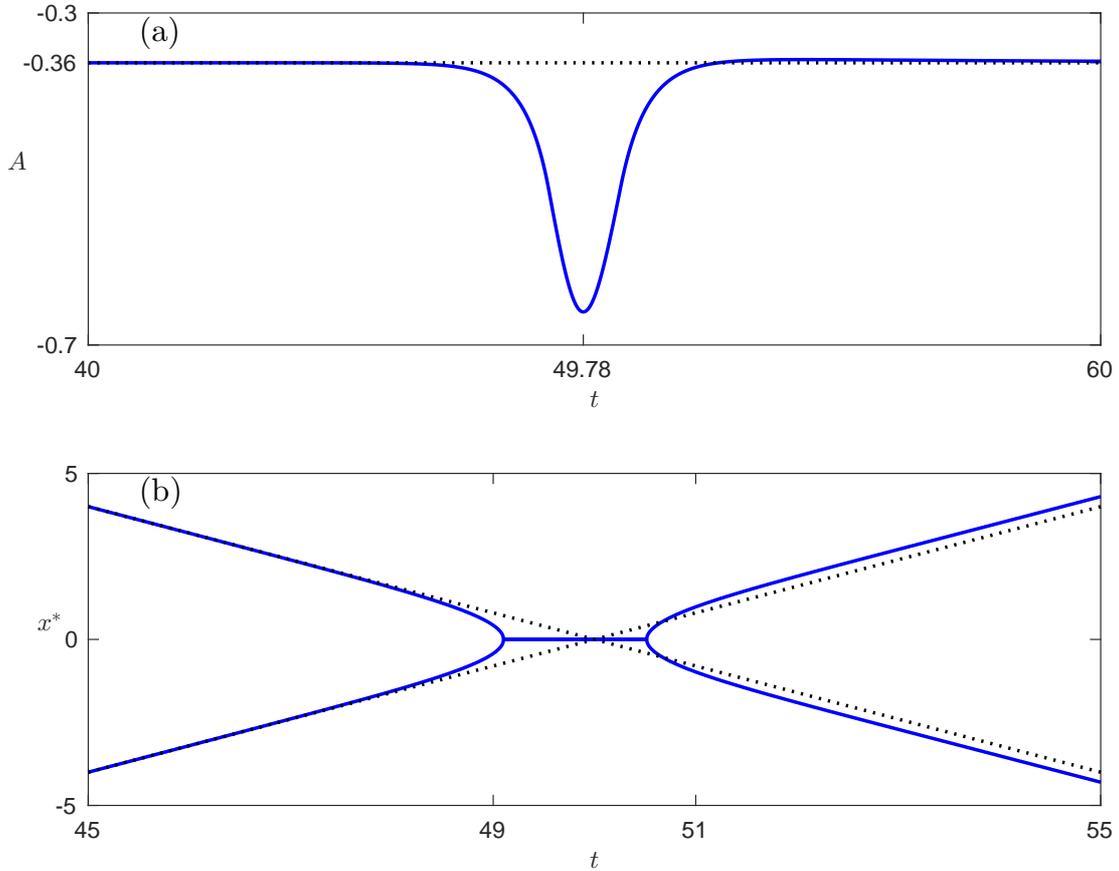}
  \caption{\small\em (a) Peak amplitude of the solution as a function of time, (b) Phase diagram of the location of the solitary waves during the interaction of Figure~\ref{fig:col3}.}
  \label{fig:col4}
\end{figure}

The minimum peak amplitude of the solution $-\,0.66$ was recorded at about $t\ =\ 49.78\,$. This value is larger than the sum of the peak amplitudes of the two solitary waves. Because of the nonlinear interaction, a phase shift can be observed in both pulses, but here the resulting solitary waves have larger amplitude $A\ \approx\ -\,0.3598$ (less negative) and therefore propagate faster than the initial solitary waves. Also, the phase change is different than the subcritical case $B\ <\ 1/3\,$. In the supercritical case $B\ >\ 1/3\,$, the waves travel faster during the interaction and are separated earlier compared to the case where small or no surface tension is considered. The minimum of the solution as a function of time and the phase diagram with the location of the peak amplitudes of the solitary waves are presented in Figure~\ref{fig:col4}. The effect of strong surface tension $B\ >\ 1/3$ on the head-on collision of solitary waves is to invert the dynamics of the collision relative to the weak surface tension case $B\ <\ 1/3\,$, resulting in faster solitary waves and dispersive tails propagating faster than the solitary waves.

\begin{figure}
  \centering
  \bigskip
  \includegraphics[width=0.99\textwidth]{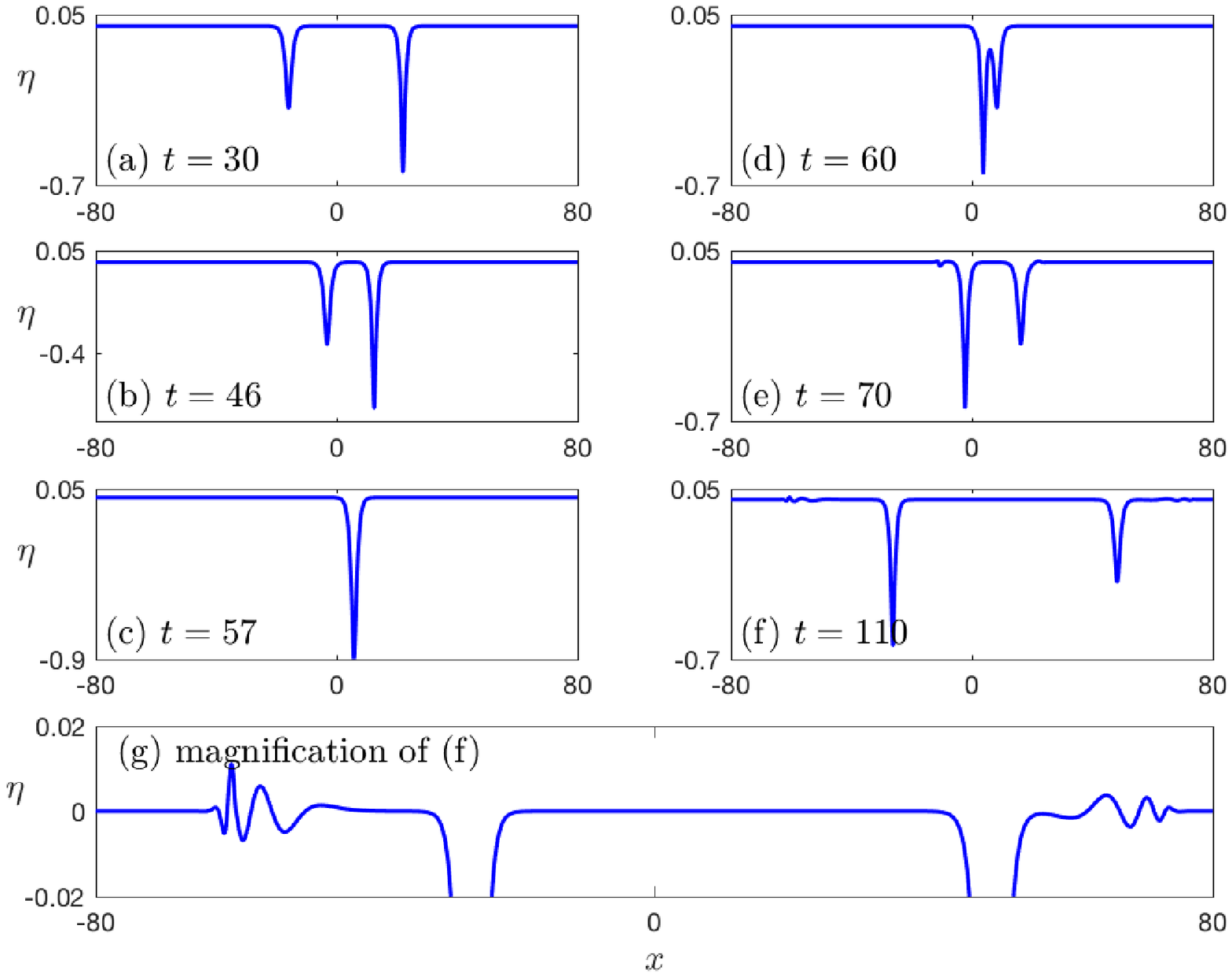}
  \caption{\small\em Unsymmetric head-on collision of two solitary waves for the \gSerre equations with $B\ =\ 0.5\,$.}
  \label{fig:col5}
\end{figure}

The collision of two solitary waves of unequal size is very similar to the one with solitary waves of equal amplitudes. For example, we consider solitary waves with speeds $c_{\,s}\ =\ 0.6$ and $0.8$ corresponding to amplitudes $A\ =\ -\,0.64$ and $-\,0.36\,$, respectively. The results of the inelastic collision are presented in Figure~\ref{fig:col5}. The dispersive tails are traveling faster than the solitary pulses again but now their shape is different, having deformed to solitary waves with different amplitudes post interaction: $A\ \approx\ -\,0.35974$ and $A\ \approx\ -\,0.63958\,$. The evolution of the minimum value of the solution is very similar with the one presented in Figure~\ref{fig:col4} and so it is omitted.


\subsection{Overtaking collisions}
\label{sec:ovc}

We now consider a different type of interaction, the overtaking collision of two solitary waves traveling in the same direction but with different speeds. In contrast to the head-on collision case, the overtaking interaction is often referred to as the \emph{strong interaction} of solitary waves due to the relative importance of nonlinearity \cite{Miles1977a}. For the \textsc{Korteweg--de Vries} (KdV) equation, a standard weakly nonlinear, long wave model of unidirectional waves, there are three categories of overtaking collisions classified by \textsc{Lax} for the two-soliton solution \cite{Lax1968}. Each category corresponds to a distinct geometry of solitary wave interaction. These categories have been observed in experiment and computation of fully nonlinear water wave models (\textsc{Euler} and \textsc{Serre} equations) \cite{CGHHS, Antonopoulos2017} and in experiment and a model of viscous core-annular flows \cite{Lowman2014}. Here, we will consider the \textsc{Lax} categories within the context of the \gSerre equations for various \textsc{Bond} numbers.

We label the three categories (\textit{a}), (\textit{b}), and (\textit{c}). \textsc{Lax} category (\textit{a}) corresponds to solitary waves of similar size whose interaction remains bimodal throughout, resulting in a small exchange of mass from the larger to the smaller solitary wave. In \textsc{Lax} category (\textit{c}), the small solitary wave is absorbed completely by the large solitary wave, resulting in a symmetric, unimodal conglomerate at the peak of interaction. Following this, the smaller solitary wave is ejected behind the larger wave and each propagates independently. \textsc{Lax} category (\textit{b}) is a combination of categories (\textit{a}) and (\textit{c}). In this case, the two solitary waves initially form an asymmetric, unimodal mass. At the peak of interaction, however, the conglomerate is bimodal. This process is undone and the smaller solitary wave is emitted behind the larger wave.

In what follows, we denote the amplitudes $A_{\,1}\,$, $A_{\,2}$ for the larger and smaller solitary waves, respectively. The amplitude ratio is denoted $r\ =\ A_{\,1}/A_{\,2}\ >\ 1\,$. For the KdV equation, interactions with $r\ \leq\ (3\ +\ \sqrt{5})/2$ are category (\textit{a}), with $(3\ +\ \sqrt{5})/2\ <\ r\ <\ 3\ \approx\ 2.6180$ category (\textit{b}) and with $r\ \geq\ 3$ category (\textit{c}).

It is important to note that the \textsc{Lax} categories for the KdV equation are completely determined by the amplitude ratio $r$ of the two solitons. This is because the KdV equation admits \textsc{Galilean} and scaling invariances that enable one to fix the leading, slower soliton to have amplitude $1$ and trailing, faster soliton to amplitude $r\,$, both on a zero background. While the \gSerre equations admit \textsc{Galilean} and scaling symmetries, the \textsc{Lax} categories for two soliton interactions functionally depend on both soliton amplitudes $A_{\,1}$ and $A_{\,2}$ separately. This is because the scaling symmetry is used to fix the total water height, here normalised to unity. In this section, we fix the amplitude of the faster \gSerre soliton to unity $A_{\,1}\ =\ 1$ and vary the slower soliton's amplitude $A_{\,2}\ <\ 1$ in order to identify the \textsc{Lax} categories in this restricted regime. Therefore, we consider the ratio $r\ =\ 1/A_{\,2}\,$, \cite{CGHHS, Lowman2014}.

In the first case with \textsc{Bond} number $B\ =\ 0.2$ similar types of interactions are observed. We use $3$ decimal digits in the calculation of the parameter $r$ and we observe that interactions with $r\ \leq\ 3.453$ are in the category (\textit{a}), for $3.469\ \leq\ r\ \leq\ 5.129$ they are in the category (\textit{b}) and for $r\ \geq\ 5.130$ they are in the category (\textit{c}). For values of $r$ in the interval $[\,3.453,\,3.468\,]\,$, we observe a transition zone between categories (\textit{a}) and (\textit{b}) where the small solitary wave is absorbed and re-emitted only after the exchange of the masses towards the end of the interaction. This phenomenon has been observed also in the case of overtaking collisions of the \textsc{Serre} equations with $3.097\ \leq\ r\ \leq\ 3.108$ \cite{Antonopoulos2017}. The limits for the three \textsc{Lax} categories for the \textsc{Serre} equations as reported in \cite{Antonopoulos2017}, along with the limits for the \gSerre (with $B\ =\ 0.2$) and \textsc{Euler} equations, \cite{CGHHS}, are summarised in Table~\ref{tab:laxlims} (for $A_{\,1}\ =\ 1$). It is noted that the values presented in Table~\ref{tab:laxlims} are correct to the digits shown and are approximate values, so they can be used as an indication of where the transition occurs.

\begin{table}
  \begin{center}
  \begin{tabular}{c|ccc}
  \hline\hline
  \textsc{Lax} \textit{categories} & (\textit{a}) &  (\textit{b}) & (\textit{c}) \\
  \hline\hline
  \textsc{Euler} & $r\ \leq\ 2.941$ & $2.941\ <\ r\ \leq\ 3.536$ & $r\ >\ 3.536$\\
  \textsc{Serre} & $r\ \leq\ 3.096$ & $3.109\ \leq\ r\ \leq\ 3.978$ & $r\ \geq\ 3.979$\\
  \gSerre & $r\ \leq\ 3.453$ & $3.469\ \leq\ r\ \leq\ 5.129$ & $r\ \geq\ 5.130$ \\
  \hline\hline
  \end{tabular}
  \end{center}
  \bigskip
  \caption{\small\em \textsc{Lax} categories for the \textsc{Euler}, \textsc{Serre} and \gSerre equations with $B\ =\ 0.2\,$.}
  \label{tab:laxlims}
\end{table}

\begin{figure}
  \centering
  \bigskip\bigskip
  \includegraphics[width=0.99\columnwidth]{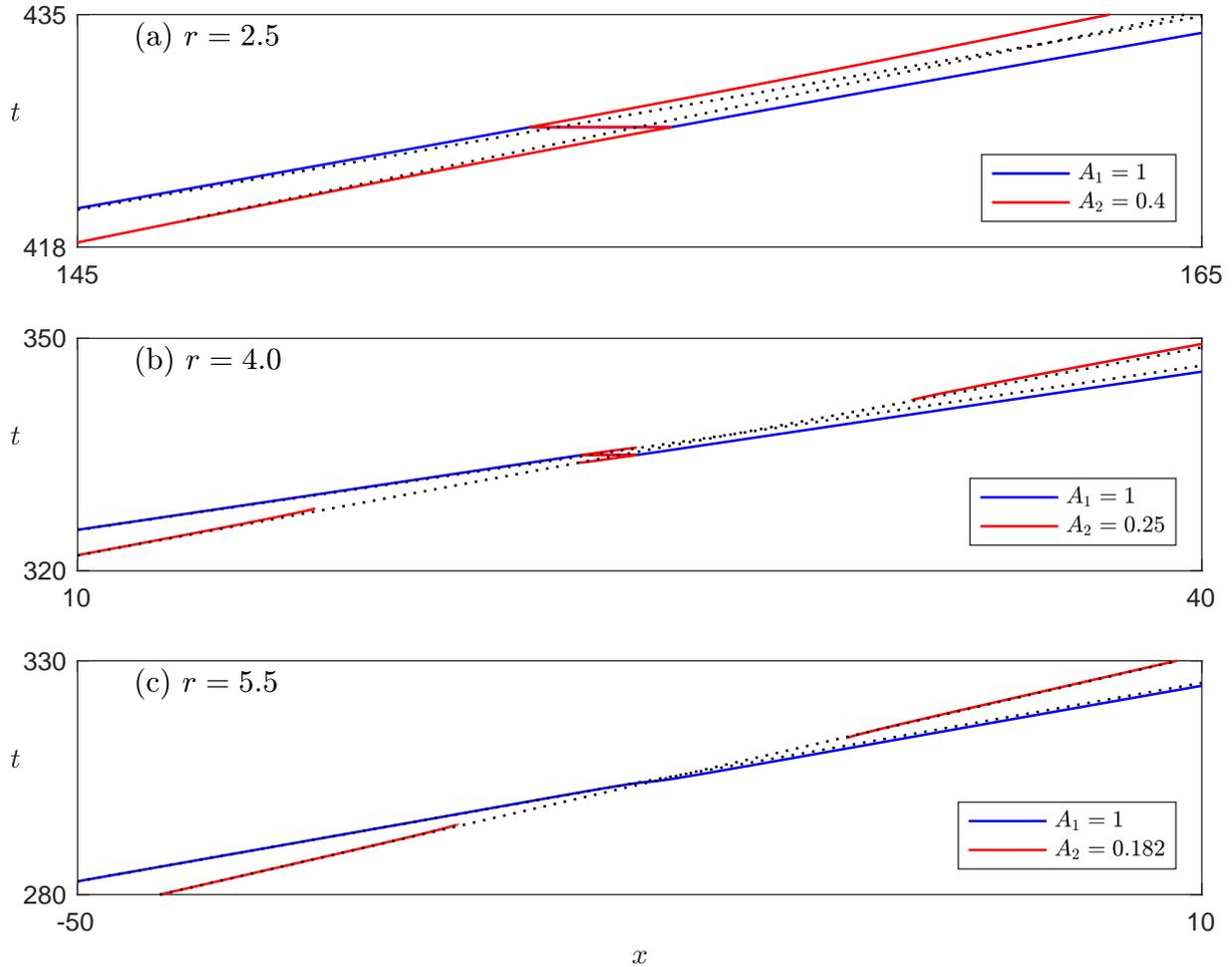}
  \caption{\small\em Phase diagrams of the three categories of \textsc{Lax} for the overtaking collision of two elevation solitary waves of the \gSerre equations with $B\ =\ 0.2\,$.}
  \label{fig:over1}
\end{figure}

Phase diagrams for the different interactions of solitary waves in each of the \textsc{Lax} categories are presented in Figure~\ref{fig:over1}. In Figure~\ref{fig:over1}(\textit{a}) it is observed that the two solitary waves maintain a distance while they exchange masses. It is also possible to observe the absorption of the small solitary wave in Figures~\ref{fig:over1}(\textit{b}) and (\textit{c}). The dotted lines represent the paths of the solitary waves as if there were no interaction.

\begin{figure}
  \centering
  \bigskip\bigskip
  \includegraphics[width=0.99\columnwidth]{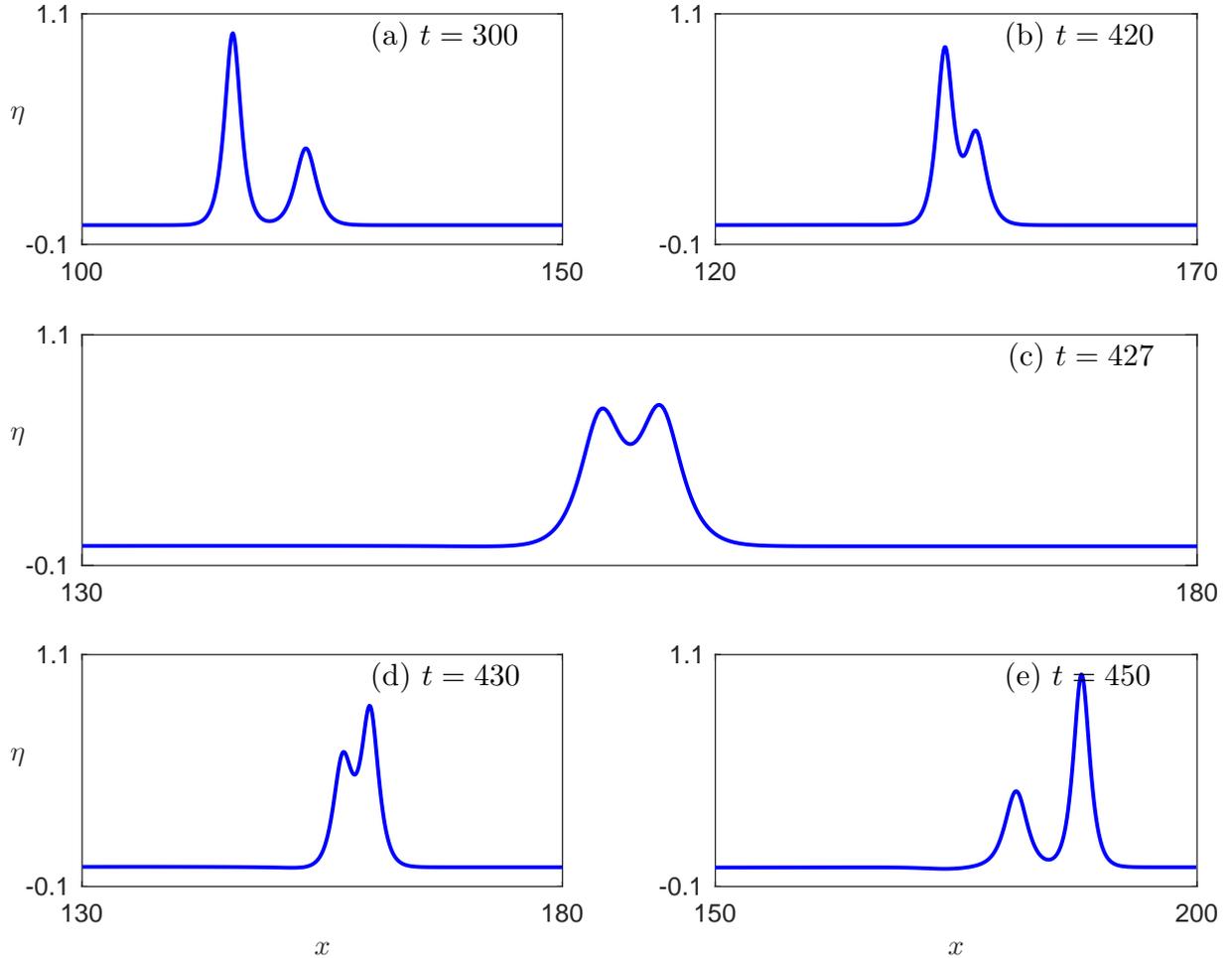}
  \caption{\small\em Interaction of two solitary waves with $r\ =\ 2.5$ of the \gSerre equations with $B\ =\ 0.2\,$. \textsc{Lax} category (a).}
  \label{fig:over2}
\end{figure}

\begin{figure}
  \centering
  \bigskip\bigskip
  \includegraphics[width=0.99\columnwidth]{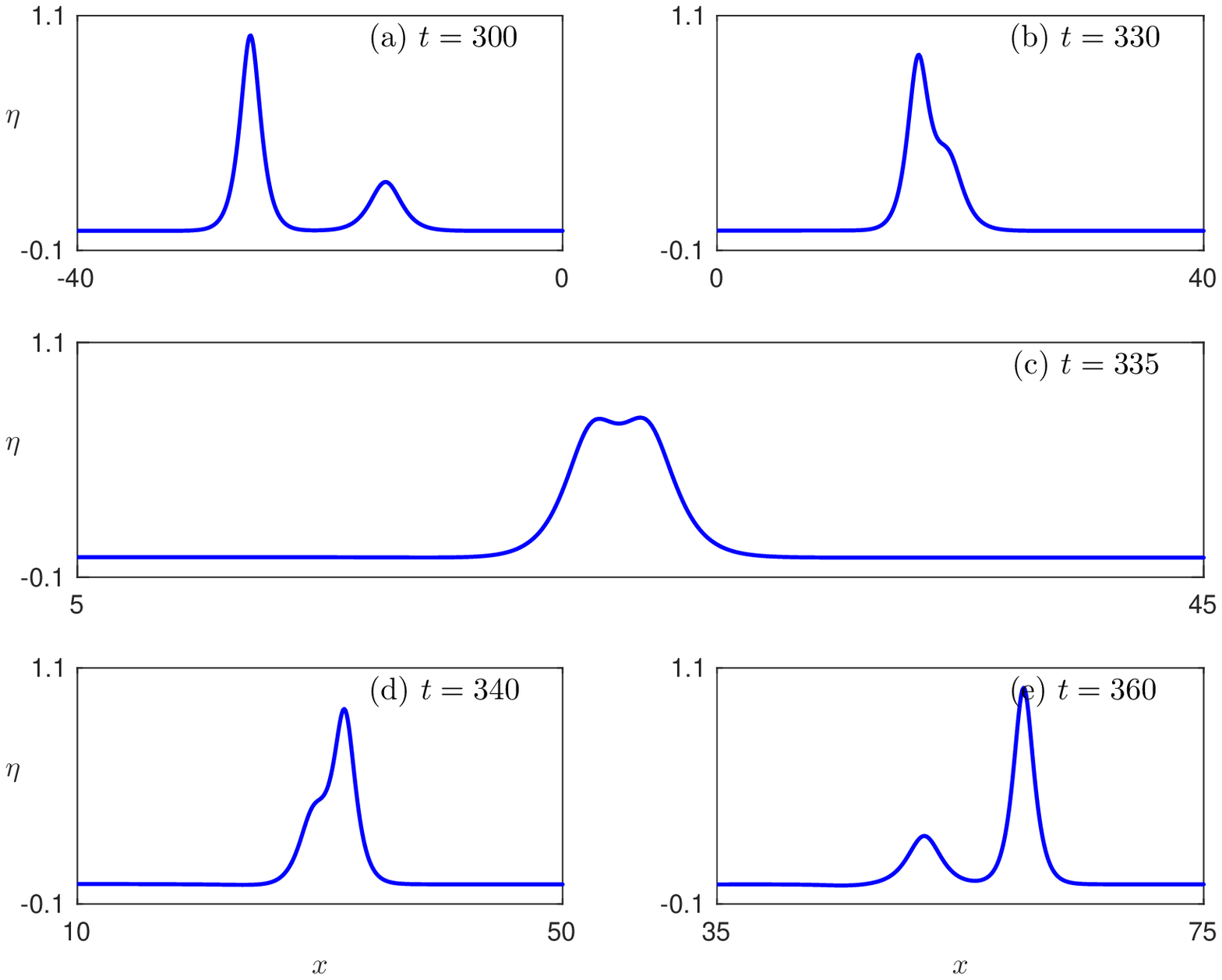}
  \caption{\small\em Interaction of two solitary waves with $r\ =\ 4$ of the \gSerre equations with $B\ =\ 0.2\,$. \textsc{Lax} category (b).}
  \label{fig:over3}
\end{figure}

\begin{figure}
  \centering
  \bigskip\bigskip
  \includegraphics[width=0.99\columnwidth]{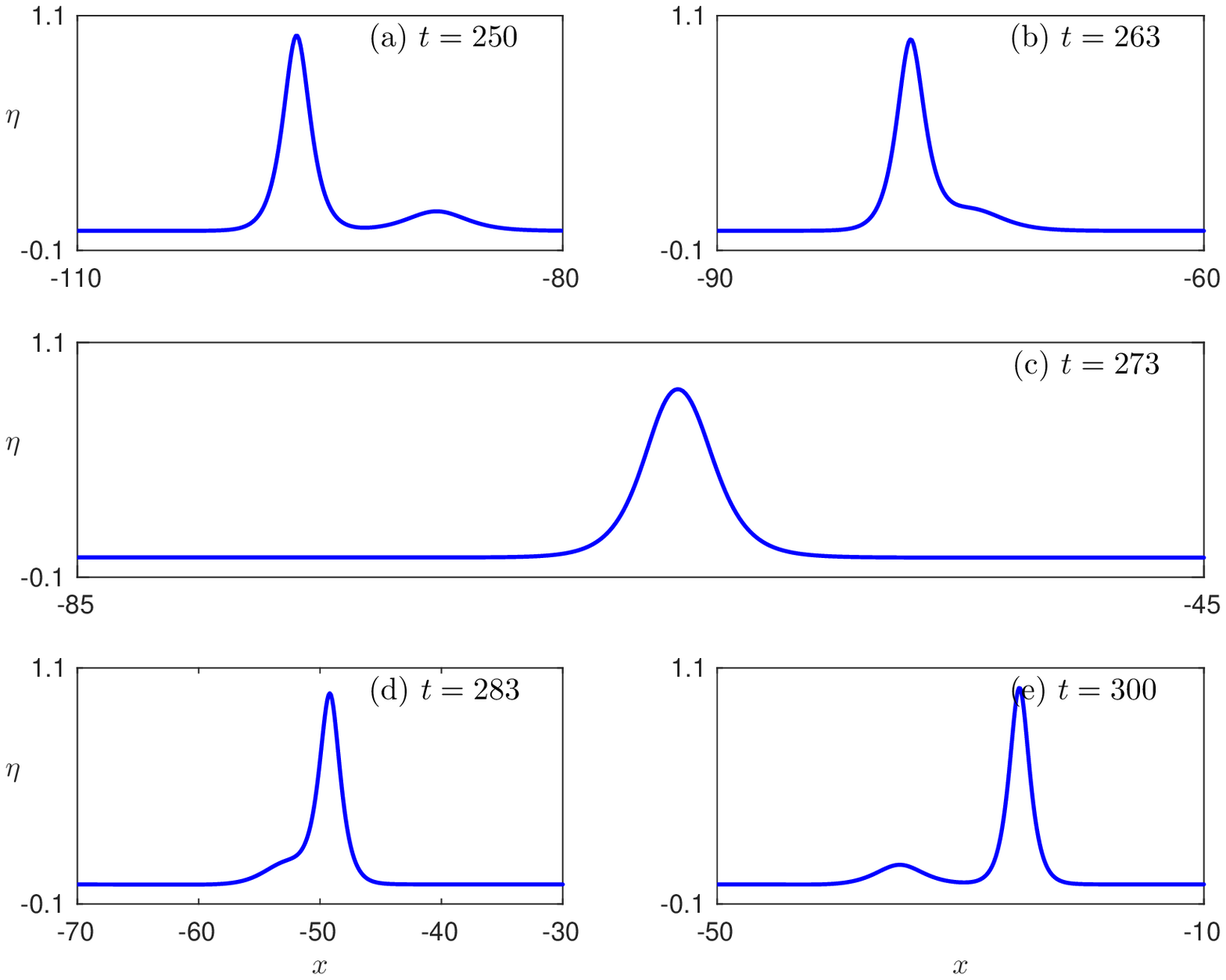}
  \caption{\small\em Interaction of two solitary waves with $r\ =\ 10$ of the \gSerre equations with $B\ =\ 0.2\,$. \textsc{Lax} category (c).}
  \label{fig:over4}
\end{figure}

The interaction with $r\ =\ 2.5$ in the \textsc{Lax} category (\textit{a}) is depicted in Figure~\ref{fig:over2}, where it is shown that during the interaction there are two peaks. In Figure~\ref{fig:over3} the interaction with $r\ =\ 4$ in the \textsc{Lax} category (\textit{b}) is presented where the small solitary wave is absorbed by the large solitary wave initially and then is re-emitted and two peaks are present during the interaction, while in the end is absorbed again and finally ejected and separated from the large solitary wave. Figure~\ref{fig:over4} shows the interaction with $r\ =\ 10\,$, which belongs to the \textsc{Lax} category (\textit{c}) where during the interaction only one peak can be observed as the small solitary wave is absorbed by the large one until it is ejected and separated from the large one at the end of the interaction.

\begin{figure}
  \centering
  \bigskip\bigskip
  \includegraphics[width=0.99\columnwidth]{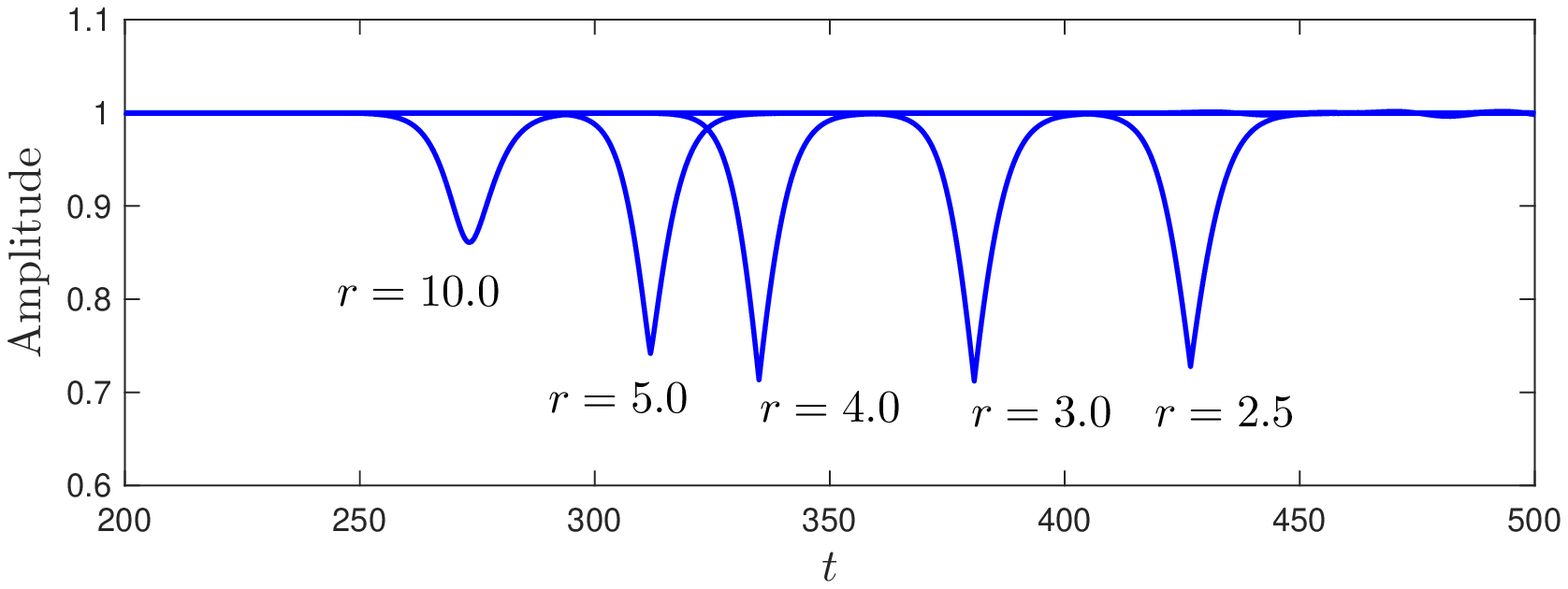}
  \caption{\small\em The amplitude of the solution for several values of $r$ for the overtaking collision of two solitary waves of the \gSerre equations with $B\ =\ 0.2\,$.}
  \label{fig:over5}
\end{figure}

The maximum value of the solution as a function of time for several values of $r$ is presented in Figure~\ref{fig:over5}. We observe that the maximum of the solution during the interaction does not behave monotonically with $r\,$. So we can achieve the same maximum for different values of $r\,$. For large values of $r$ and in the category (\textit{c}) of \textsc{Lax}, the amplitude as a function of time is a smooth function. In the categories (\textit{a}) and (\textit{b}) the maximum appeared to have a singularity at the time $t$ where the minimum occurs.

Although the limits for the \textsc{Lax} categories of overtaking collisions are different when surface tension is included (\cf, Table~\ref{tab:laxlims}), the interactions are very similar and no new phenomena are observed when compared to the results reported in \cite{Mirie1982, CGHHS, Li2004, Antonopoulos2017, Mitsotakis2014}.

\begin{figure}
  \centering
  \bigskip\bigskip
  \includegraphics[width=0.99\columnwidth]{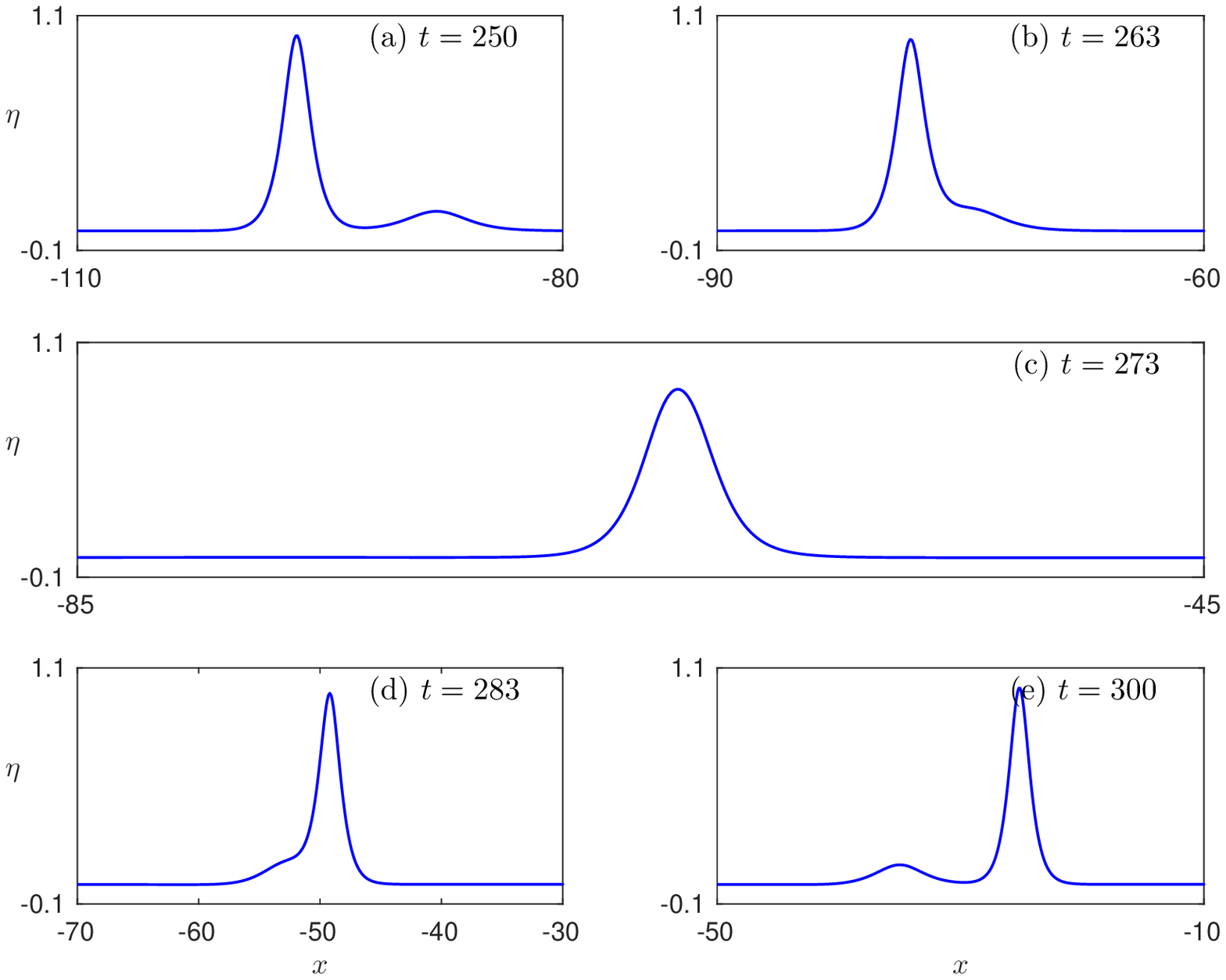}
  \caption{\small\em Dispersive tails generated during the overtaking collision of two solitary waves of the \gSerre equations with $B\ =\ 0.2\,$, $r\ =\ 4\,$.}
  \label{fig:over6}
\end{figure}

Like in the case of the \textsc{Serre} equations, small amplitude dispersive tails are generated during and after the interaction of two solitary waves. The dispersive tails are propagating mainly to the right but a small $N-$shaped wavelet is generated and propagates to the left as shown in Figure~\ref{fig:over6}.

We draw the conclusion that the overtaking collision of two solitary waves for the \gSerre equations with $B\ <\ 1/3$ retains the qualitative characteristics of the analogous interaction for the \textsc{Serre} equations. The situation for the \gSerre equations with $B\ >\ 1/3$ is somewhat different, as we now demonstrate. We consider the \gSerre equations with $B\ =\ 0.5$ and we test overtaking collisions for different solitary waves. In this case, we were able to observe the analogous three categories of \textsc{Lax} but here, the values of $r$ are totally different and cannot be compared with the analogous results obtained for small values of \textsc{Bond} number $B\,$. Specifically, we define here $r\ =\ \abs{a_{\,2}}/\abs{a_{\,1}}$ with $\abs{a_{\,1}}\ >\ \abs{a_{\,2}}$ \ie the amplitude of the small solitary wave over the amplitude of the large solitary wave.

In order to compare numerical results with the predictions of \textsc{Lax}, we restrict our attention to solitary waves of small amplitude where the unidirectional solitary waves of the \gSerre equations can be asymptotically approximated by KdV solitary waves. All the experiments were performed in $[\,-200,\,200\,]$ and the solitary waves are translated initially so as to attain their maximum values at $x\ =\ -\,50$ and $x\ =\ 50$ respectively. When $B\ <\ 1/3$ we take $\Delta x\ =\ 0.1$ and $\Delta t\ =\ 0.01$ while, for $B\ >\ 1/3$ we take $\Delta x\ =\ 0.02$ and $\Delta t\ =\ 0.01\,$.

\begin{figure}
  \centering
  \bigskip\bigskip
  \includegraphics[width=0.99\columnwidth]{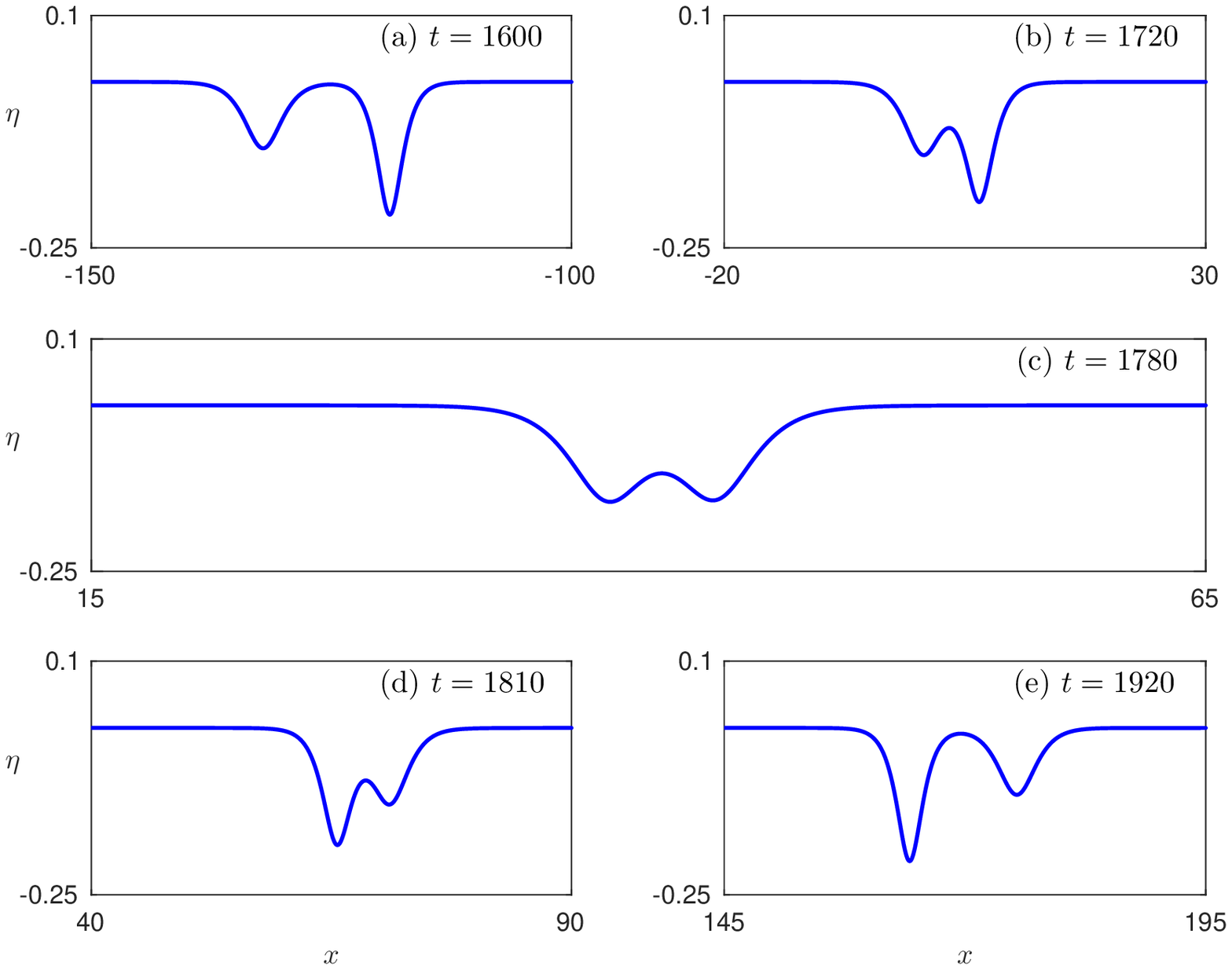}
  \caption{\small\em Interaction of two solitary waves with $r\ =\ 2$ of the \gSerre equations with $B\ =\ 0.5\,$. \textsc{Lax} category (a).}
  \label{fig:over7}
\end{figure}

In these experiments, the fast solitary wave has amplitude $a_{\,1}\ =\ -0.1\,$. Then, for a small solitary wave of amplitude $a_{\,2}\ =\ -\,0.2\,$, the interaction falls into \textsc{Lax} category (\textit{a}). As can be observed in Figure~\ref{fig:over7}, the solitary waves exchange masses and there are always two pulses present in the domain. On the other hand, when we take $a_{\,2}\ =\ -\,0.3\,$, the small solitary wave is initially absorbed by the large solitary wave, and then re-emitted, causing the existence of two local minima during the interaction, as described by \textsc{Lax} category (\textit{b}). Finally, for $a_{\,2}\ =\ -\,0.4$ the small solitary wave is absorbed by the large one during the interaction and after the interaction is ejected and separated from the large one, \textsc{Lax} category (\textit{c}).

\begin{figure}
  \centering
  \bigskip\bigskip
  \includegraphics[width=0.99\columnwidth]{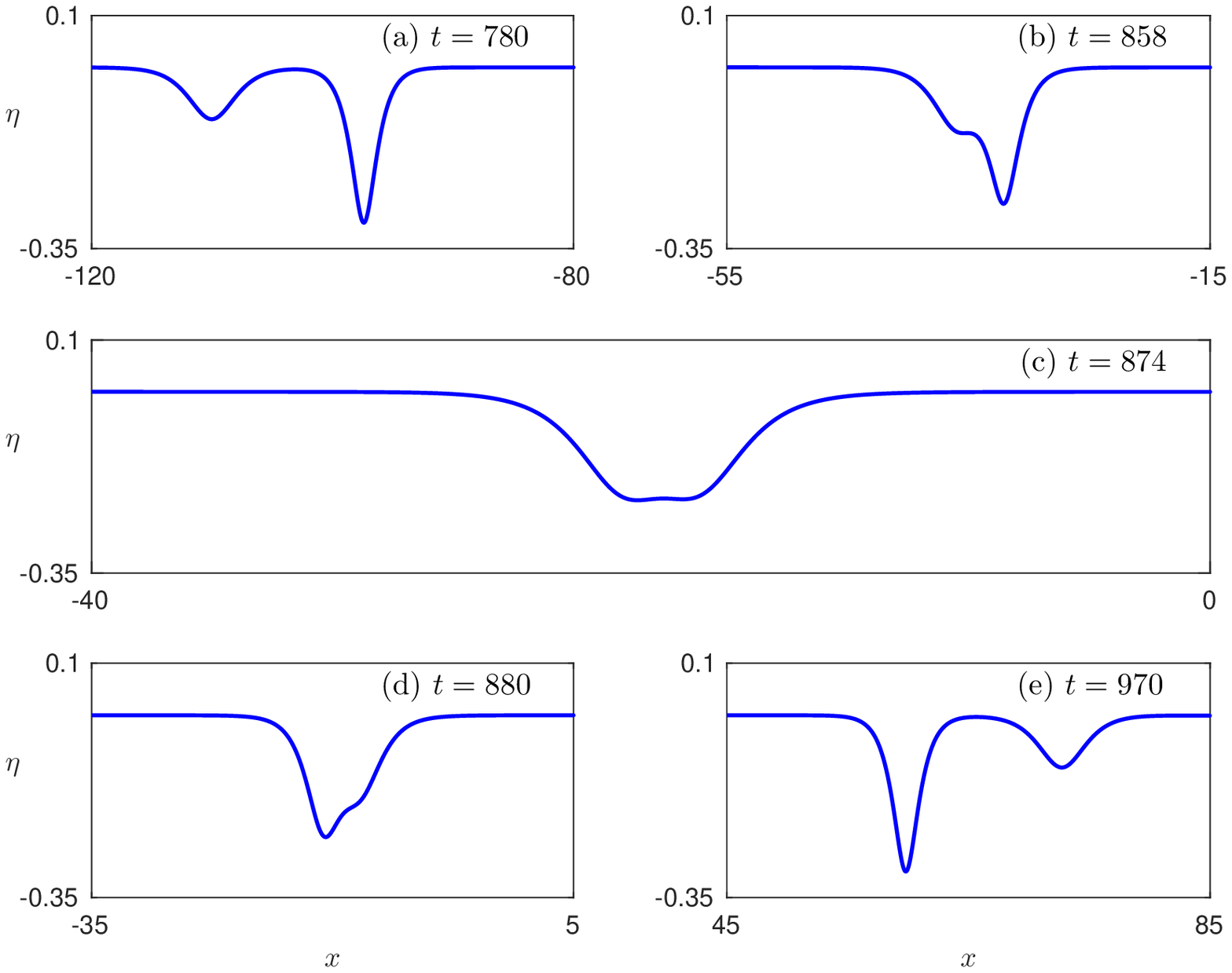}
  \caption{\small\em Interaction of two solitary waves with $r\ =\ 3$ of the \gSerre equations with $B\ =\ 0.5\,$. \textsc{Lax} category (b).}
  \label{fig:over8}
\end{figure}

\begin{figure}
  \centering
  \bigskip
  \includegraphics[width=0.9\columnwidth]{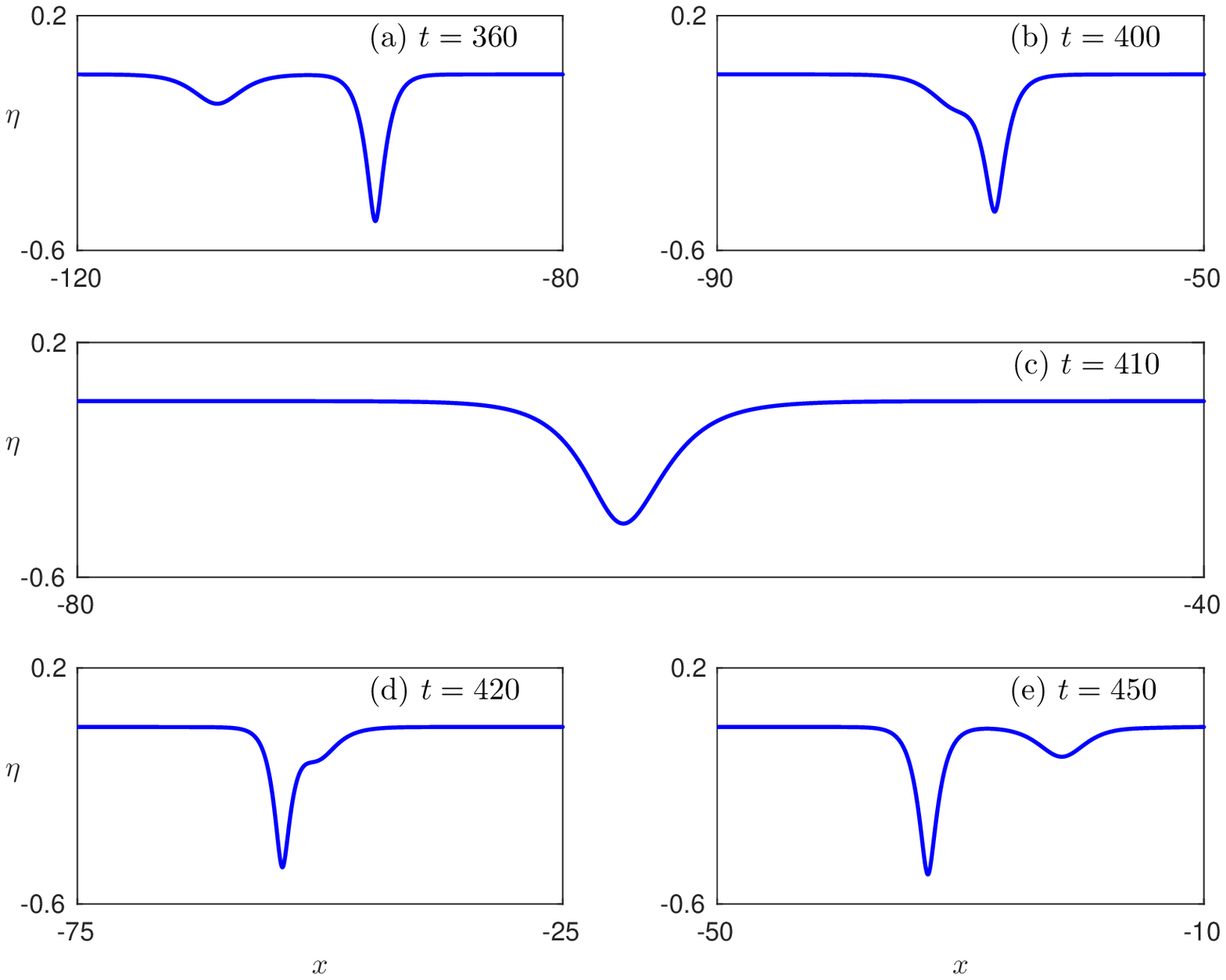}
  \caption{\small\em Interaction of two solitary waves with $r\ =\ 10$ of the \gSerre equations with $B\ =\ 0.5\,$. \textsc{Lax} category (c).}
  \label{fig:over9}
\end{figure}

\begin{figure}
  \centering
  \bigskip\bigskip
  \includegraphics[width=0.99\columnwidth]{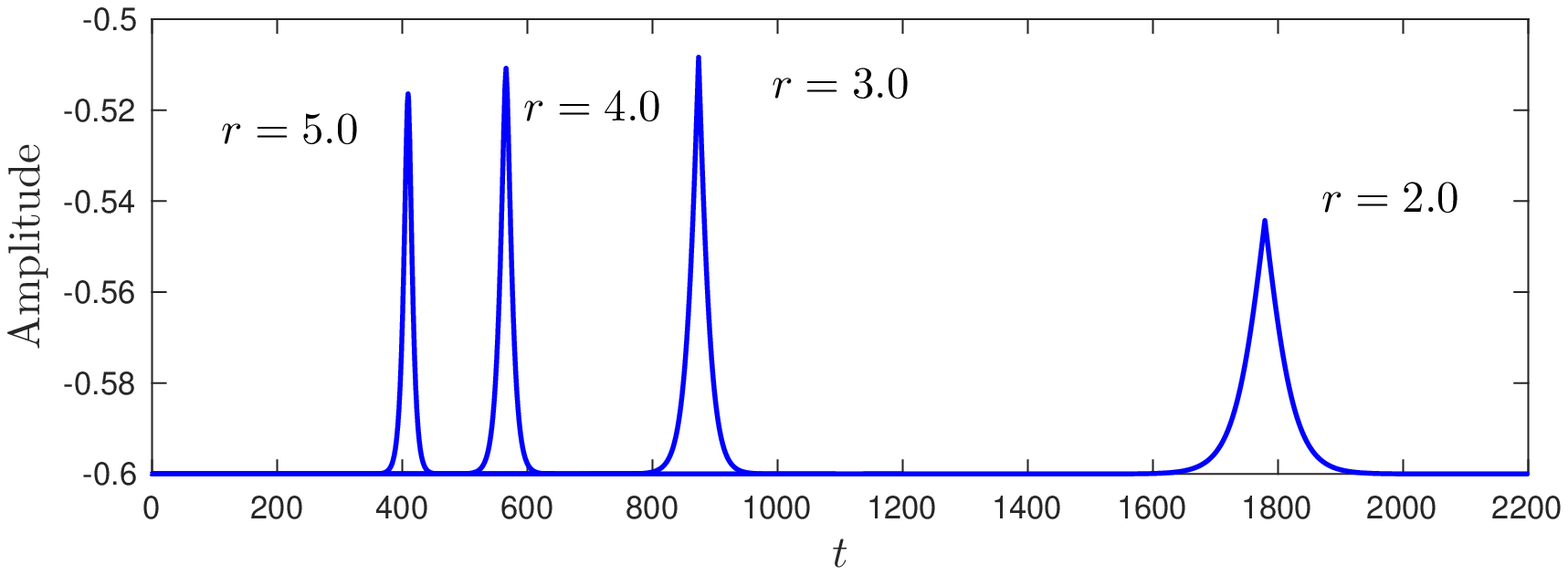}
  \caption{\small\em The maximum negative amplitude of the solution for several values of $r$ for the overtaking collision of two solitary waves of the \gSerre equations with $B\ =\ 0.5\,$.}
  \label{fig:over10}
\end{figure}

Again, the minimum value of the solution as a function of time behaves similarly to the case of small \textsc{Bond} number but here the amplitude is the negative minimum of the solution. Figure~\ref{fig:over10} shows the evolution of the amplitude as a function of time for different values of $r\,$. In this figure, the amplitudes have been translated so as to be all at the same level. Again, we observe that as the interaction changes categories, the maximum amplitude is not a monotone function with respect to $r\,$.

\begin{figure}
  \centering
  \bigskip
  \includegraphics[width=0.9\columnwidth]{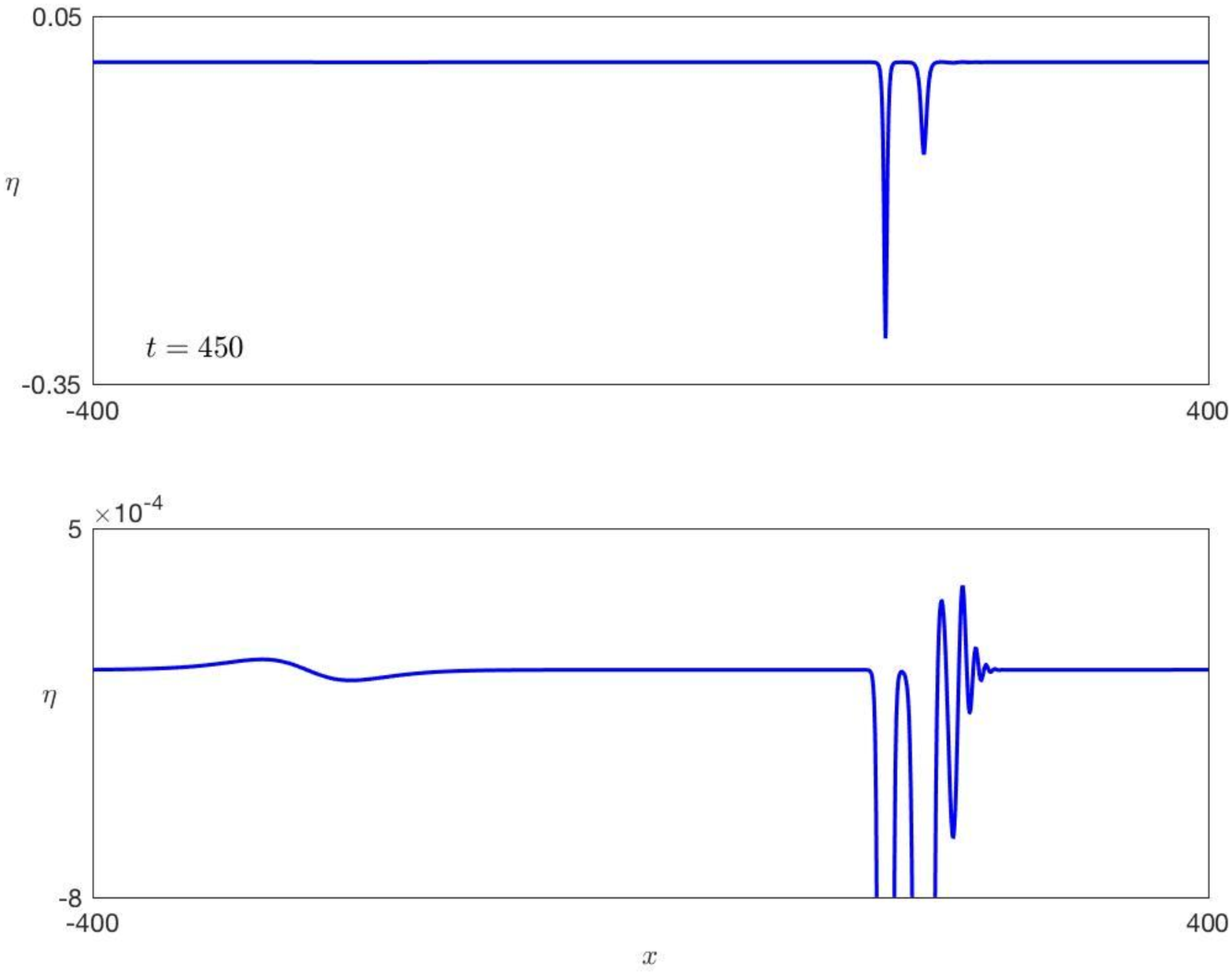}
  \caption{\small\em Dispersive tails generated during the overtaking collision of two solitary waves of the \gSerre equations with $B\ =\ 0.5\,$, $r\ =\ 3\,$.}
  \label{fig:over11}
\end{figure}

The interaction again is inelastic and the generation of small amplitude dispersive tails is observed. Figure~\ref{fig:over11} shows the dispersive tails generated by the interaction of two solitary waves with $a_{\,1}\ =\ -\,0.1$ and $a_{\,2}\ =\ -\,0.3\,$. Again, a small amplitude oscillatory dispersive tail is generated in front of the two pulses that travels faster than the pulses. On the other hand, a very fast $N-$shaped wavelet is also generated moving to the left. The tails generated in the other cases where $B\ >\ 1/3$ are always very similar, thus we do not present them here.

\subsection{Critical and transcritical cases}
\label{sec:crit}

In this section, we study the critical case of \textsc{Bond} number $B\ =\ 1/3$ and the cases where $B$ is close to this critical value. As it was mentioned in Section~\ref{sec:numeth}, and in \cite{Dias2010}, there are no smooth traveling wave solutions known for the critical \textsc{Bond} number $B\ =\ 1/3$ but only the peaked solitary waves given by the formula \eqref{eq:peakon}. Additionally, the solitary waves corresponding to \textsc{Bond} numbers close to the critical value are very close to peaked solitary waves. Moreover, the absence of dispersive effects for the critical value $B\ =\ 1/3$ is expected to influence the behaviour of the solutions for values of $B$ close to $1/3$ is expected to be affected by. We first study the interactions of solitary waves in the cases where $B\ =\ 0.32\,, 0.33\,, 0.34$ and $0.35\,$.

\begin{figure}
  \centering
  \bigskip
  \includegraphics[width=0.9\columnwidth]{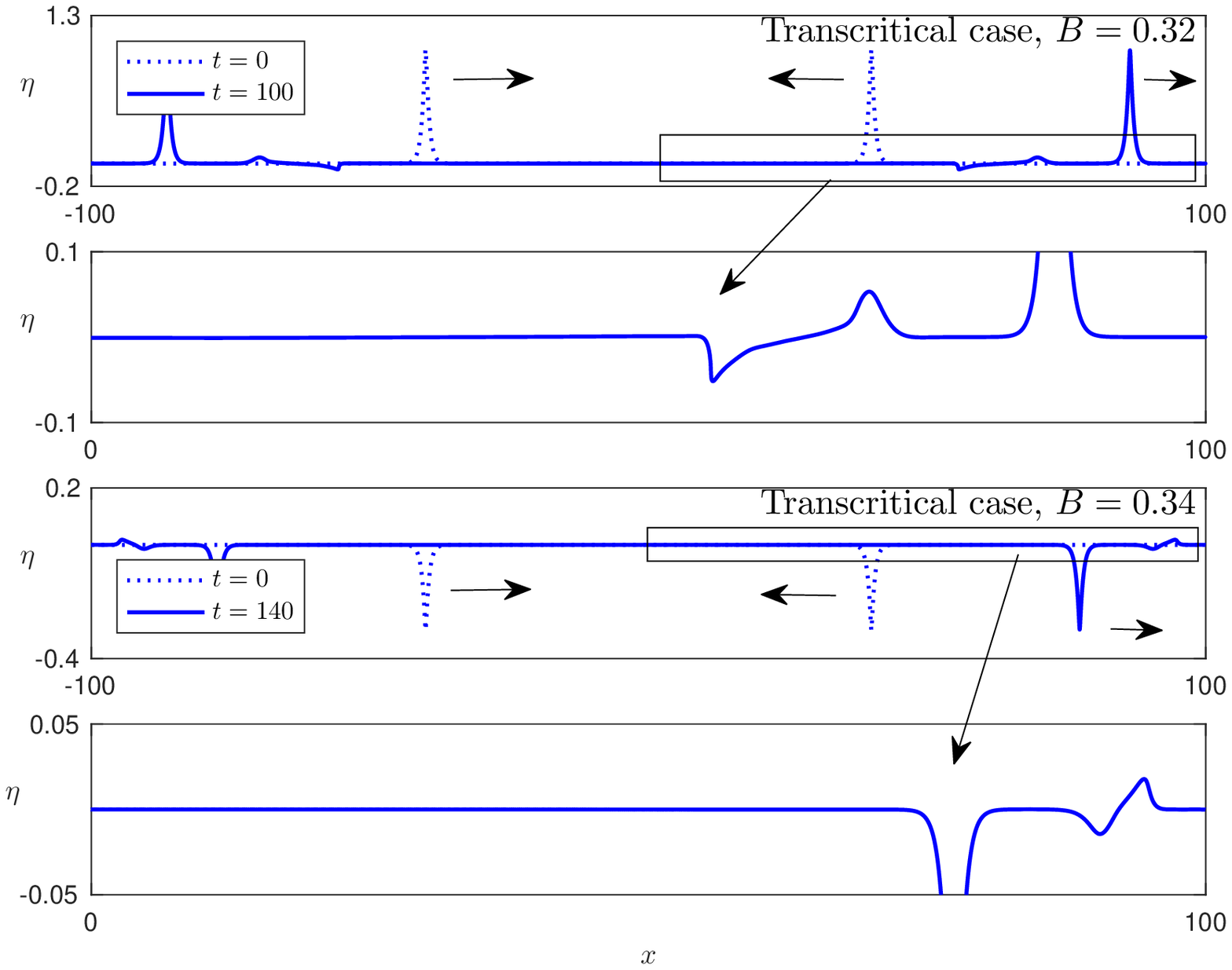}
  \caption{\small\em Head-on collisions of solitary waves of equal amplitude for the transcritical values of \textsc{Bond} number $B\ =\ 0.32$ and $B\ =\ 0.34\,$.}
  \label{fig:critical2}
\end{figure}

We start with the description of the head-on collision of two equal solitary waves. For subcritical values of the \textsc{Bond} number, we consider solitary waves of amplitude $A\ =\ 1\,$, while we take $A\ =\ -\,0.3$ for the supercritical values. Although the two waves interact in a similar manner with the cases described in Section~\ref{sec:numex}, the generated dispersive tails are not oscillatory; rather, they are similar to $N-$shaped waves. For example, when $B\ =\ 0.34\,$, the dispersive tail is ahead of the solitary wave, while in the case $B\ =\ 0.33$ the tails are behind the solitary waves. A close look at the dispersive tails shows an intermediate state between dispersive and non-dispersive waves. Figure~\ref{fig:critical2} presents the results after the interaction of two equal solitary waves for $B\ =\ 0.32$ and $B\ =\ 0.34\,$.

\begin{figure}
  \centering
  \bigskip
  \includegraphics[width=0.9\columnwidth]{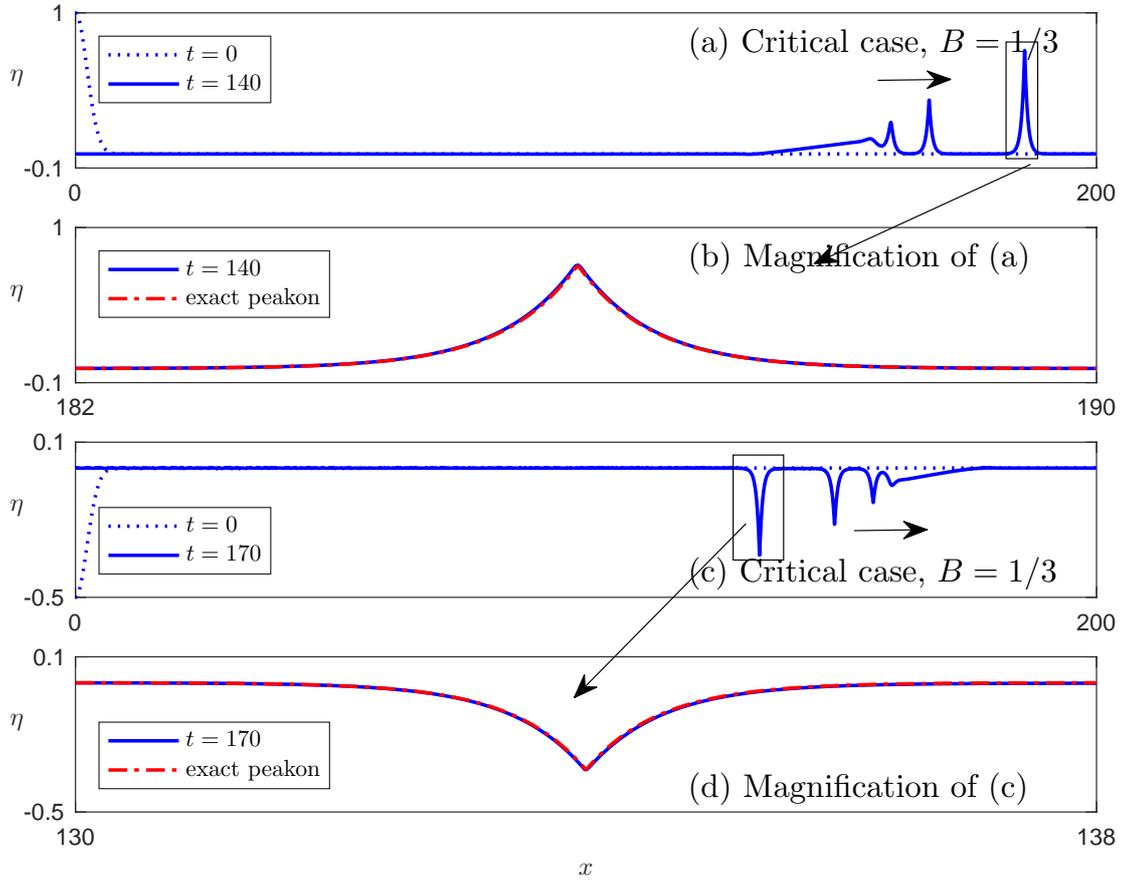}
  \caption{\small\em Evolution of \textsc{Gaussian} initial conditions into a series of elevation and depression solitary waves for the critical \textsc{Bond} number $B\ =\ 1/3\,$.}
  \label{fig:critical1}
\end{figure}

\begin{figure}
  \centering
  \bigskip
  \includegraphics[width=0.99\columnwidth]{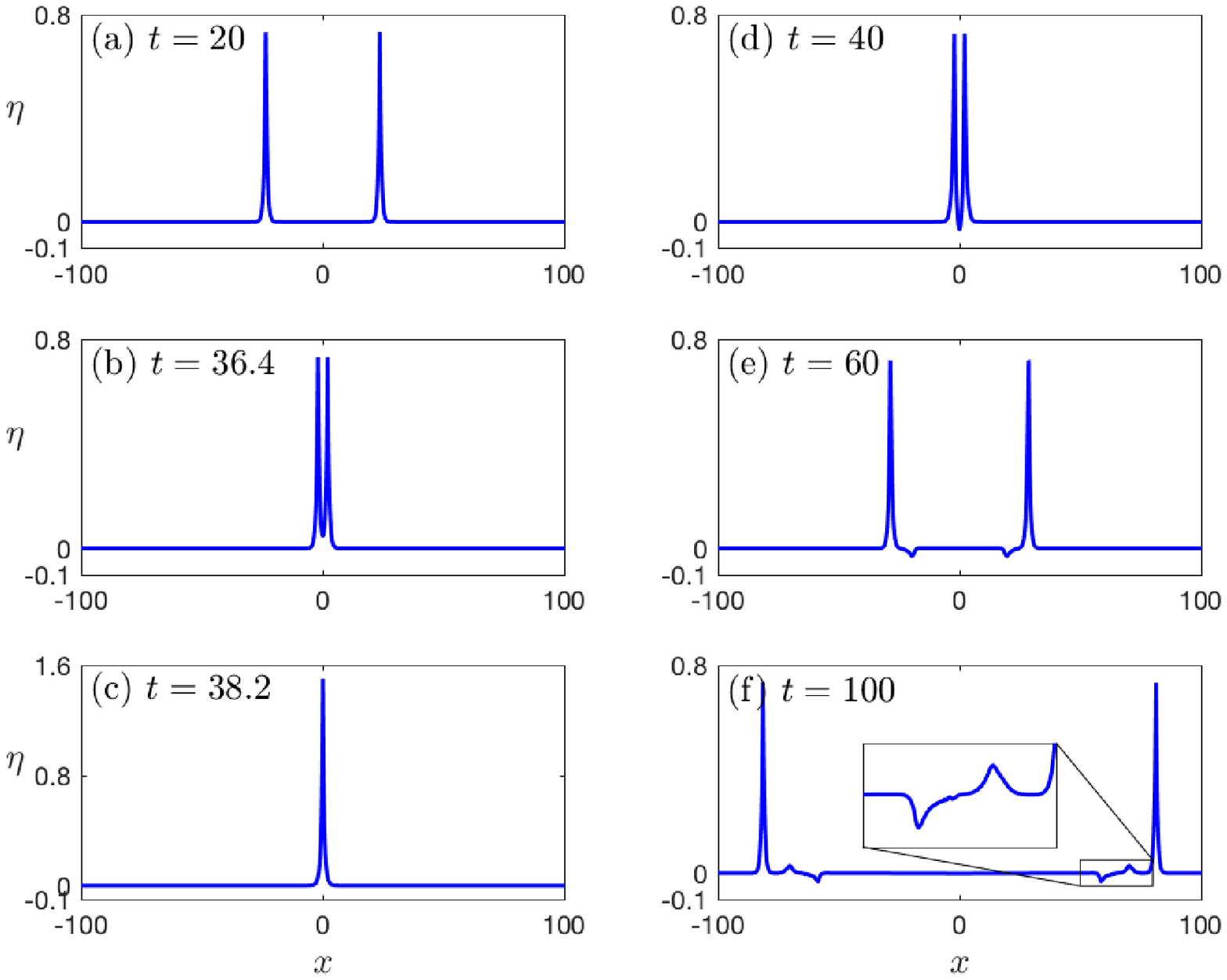}
  \caption{\small\em Head-on collision of cleaned peakons for the critical \textsc{Bond} number $B\ =\ 1/3\,$.}
  \label{fig:hocrit}
\end{figure}

The interactions of solitary waves in all transcritical cases we tested are very similar to the analogous subcritical and supercritical cases. The main difference observed between the two cases is the shape of the dispersive tails. Analogous behavior is observed for the overtaking collision in the transcritical cases. The overtaking collisions generated small $N-$shaped wavelets and dispersive tails, which consist of a few oscillations and are very similar to the tails generated after the head-on collision. For this reason, we do not present pictures of the overtaking collisions in the transcritical cases in this paper. We proceed with the critical case $B\ =\ 1/3\,$.

Although the critical case is numerically difficult in the sense that numerical errors might lead to false conclusions, we explore the existence of stable solitary waves when $B\ =\ 1/3\,$. It is known that localized initial conditions evolve into a series of solitary waves and dispersive tails in many nonlinear and dispersive wave models, \cite{Dias2010, BC}. In order to examine if the \gSerre equations possess stable solitary wave solutions when $B\ =\ 1/3\,$, we test the evolution of a general initial condition of the form $\eta\,(x,\,0)\ =\ a\,\ue^{\,-\,b\,x^{\,2}}$ with zero initial velocity $u\,(x,\,0)\ =\ 0\,$. We present the results for $a\ =\ 1$ and $b\ =\ 0.1\,$, \ie when we consider the evolution of a heap of water under gravity. The initial waveform is split into two symmetric waves that eventually evolve into a series of solitary waves. In Figures~\ref{fig:critical1}(\textit{a}) and (\textit{b}), we present the resolution of a \textsc{Gaussian} initial condition into a series of solitary waves where we set $\Delta x\ =\ 0.02$ and $\Delta t\ =\ 0.002\,$. Although the computation is performed in the interval $[\,-200,\,200\,]\,$, we present only the solution in the interval $[\,0,\,200\,]$ because it is symmetric. In Figure~\ref{fig:critical1}(\textit{b}), we observe that the solitary waves coincide with peaked solitary waves when we compare the shape of the numerical solution with the analogous analytical peakon \eqref{eq:peakon}. This example serves as an indication of the existence of stable peaked solitary waves for the \gSerre equations in the critical case $B\ =\ 1/3\,$.

The situation is similar when we take a negative initial condition given by the same formula with $a\ =\ -0.5\,$, $b\ =\ 0.1$ for the critical value $B\ =\ 1/3\,$. It is observed that although the initial condition is negative, it is resolved into a series of depression, peaked solitary waves (also known as antipeakons), indicating that the \gSerre equations possess both stable depression and elevation peaked solitary waves when $B\ =\ 1/3\,$. In this experiment, we use $\Delta x\ =\ 10^{\,-3}$ and $\Delta t\ =\ 10^{\,-4}$ and the interval of integration is $[\,-200,\,200\,]\,$.  Although further theoretical studies are required to ensure the accuracy of our conclusions, some confidence in the numerical results can be gained by the fact that both elevation and depression traveling waves appear to exist at the same time when $B\ =\ 1/3\,$, contrary to what we have experienced for large and small values of \textsc{Bond} number $B\ \neq\ 1/3\,$, where we are able to compute either elevation or depression solitary waves in each case.

In order to study the solitary waves generated by the evolution of a \textsc{Gaussian} and to ensure that they are \emph{bona fide} peakons in this critical case, we isolated the solitary waves using the cleaning procedure suggested in \cite{DDMM, BC}. After one cleaning iteration, the solitary wave propagates without change in shape, amplitude or speed. Specifically, the elevation solitary wave of Figure~\ref{fig:critical1}(\textit{a}) propagates with constant speed $c_{\,s}\ \approx\ 1.317$ and amplitude $A\ \approx\ 0.7337$ and satisfies the usual speed-amplitude relationship. The amplitude of the emerging depression wave is approximately $A\ \approx\ -0.3368\,$. The interaction of the cleaned solutions are studied and found to be very similar to the transcritical cases. The symmetric head-on collision of two elevation peakons is presented in Figure~\ref{fig:hocrit} where we observe similarities with the head-on collision of two solitary waves in the case $B\ =\ 0.32\,$. Similarly, the head-on collision of two depression peakons leads to very similar results when compared to the case $B\ =\ 0.34\,$. This suggests that at least some dynamics of the \gSerre equations with $B$ near-critical are reasonably approximated by dynamics of the critical case.

\begin{figure}
  \centering
  \bigskip
  \includegraphics[width=0.9\columnwidth]{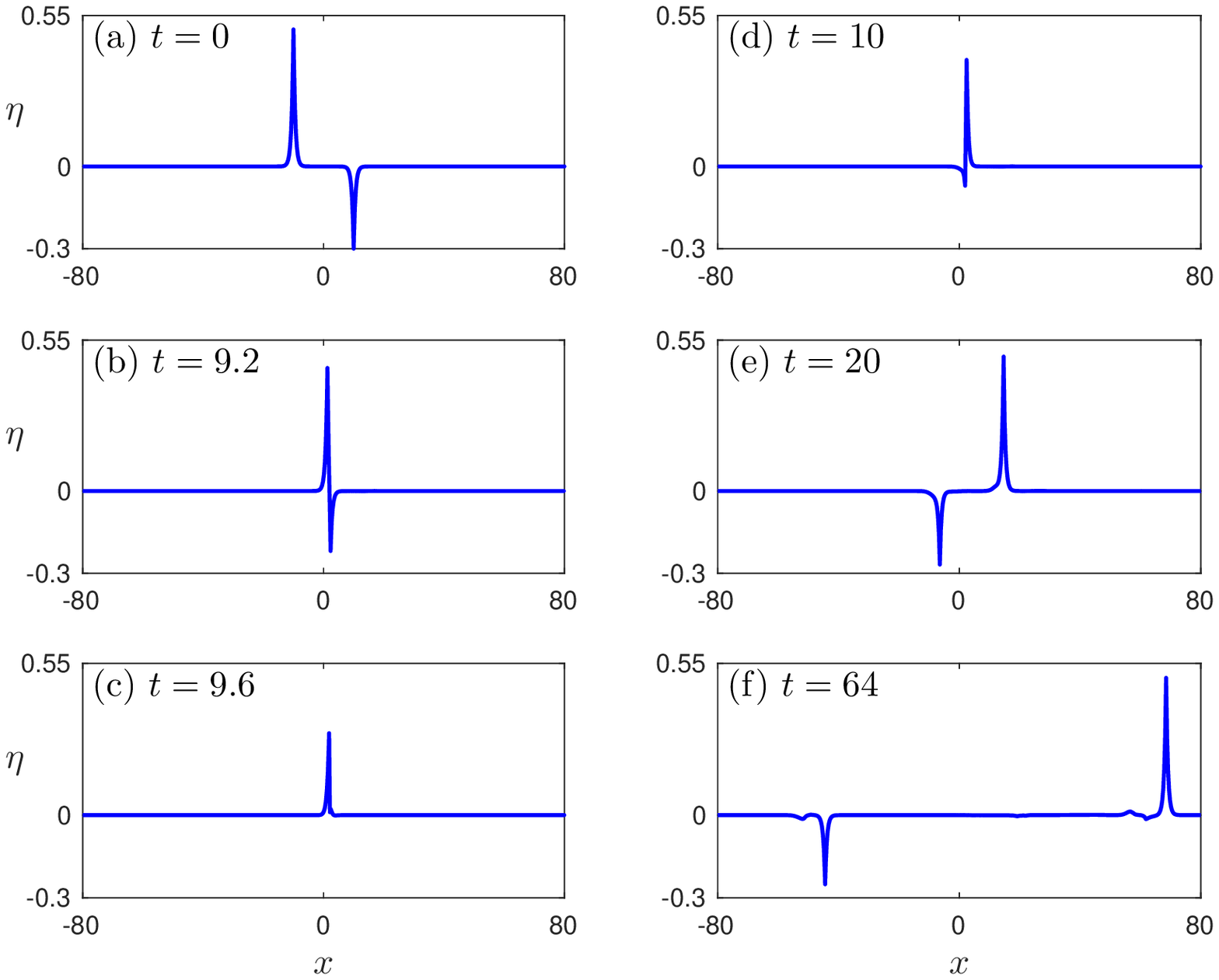}
  \caption{\small\em Head-on collision between elevation and depression peakons for the critical \textsc{Bond} number $B\ =\ 1/3\,$.}
  \label{fig:mixedcol}
\end{figure}

\begin{figure}
  \centering
  \bigskip
  \includegraphics[width=0.9\columnwidth]{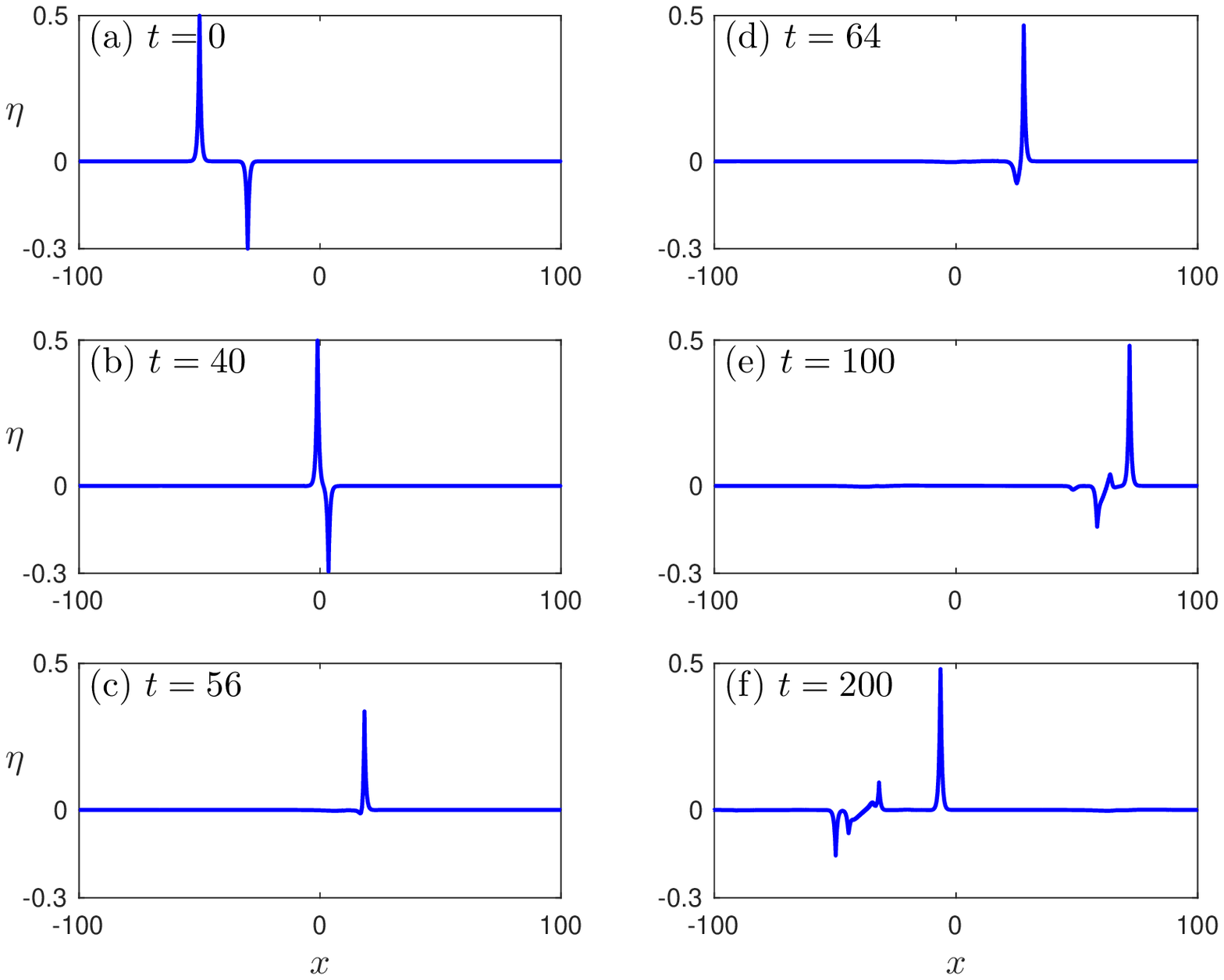}
  \caption{Overtaking collision between elevation and depression peakons for the critical \textsc{Bond} number $B\ =\ 1/3\,$.}
  \label{fig:mixhead}
\end{figure}

\begin{figure}
  \centering
  \bigskip
  \includegraphics[width=0.9\columnwidth]{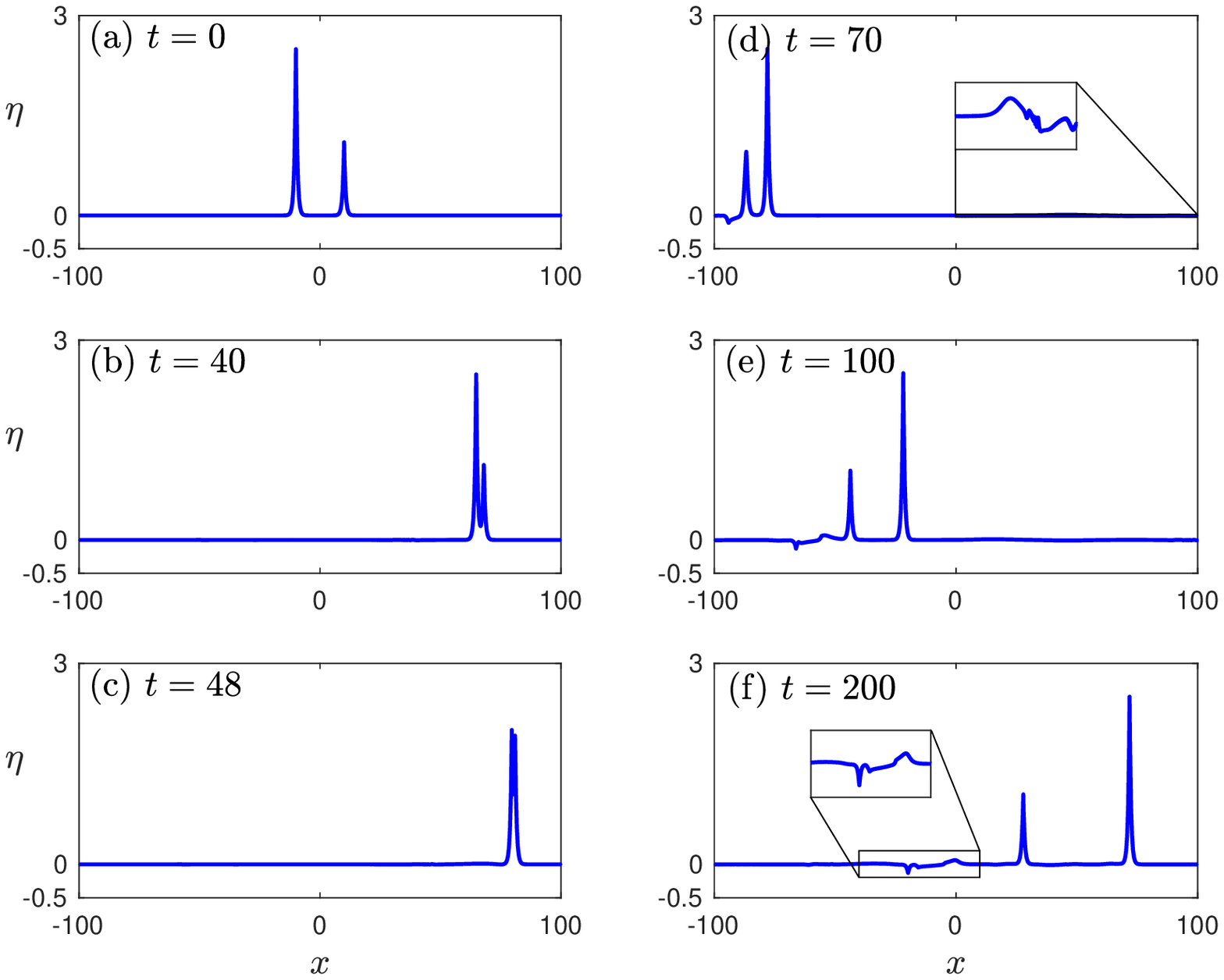}
  \caption{\small\em Overtaking collision between elevation peakons for the critical \textsc{Bond} number $B\ =\ 1/3\,$.}
  \label{fig:ovpeak1}
\end{figure}

\begin{figure}
  \centering
  \bigskip
  \includegraphics[width=0.9\columnwidth]{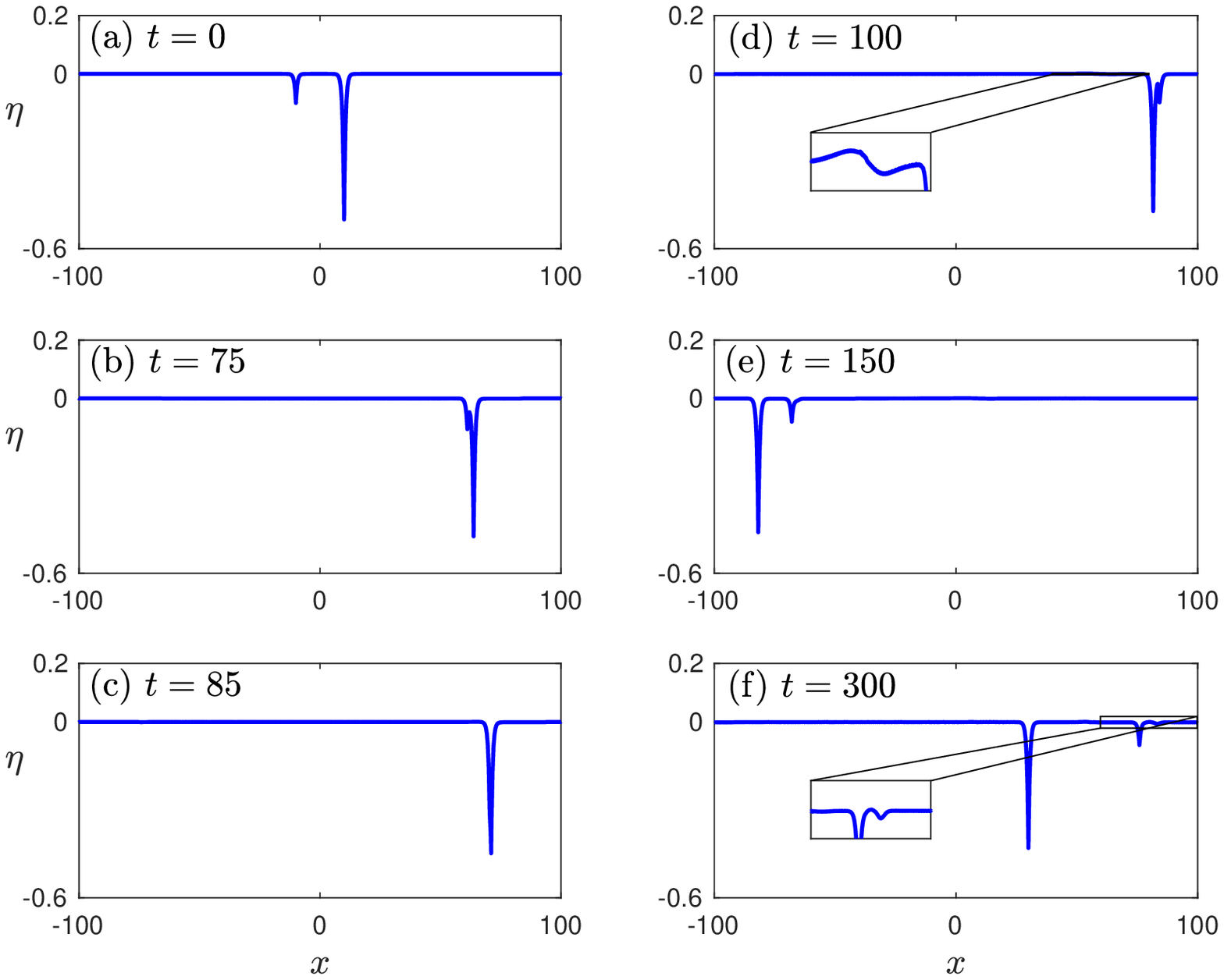}
  \caption{\small\em Overtaking collision between depression peakons for the critical \textsc{Bond} number $B\ =\ 1/3\,$.}
  \label{fig:ovpeak2}
\end{figure}

We also explore the head-on collision between a peakon of elevation and a peakon of depression with amplitudes $A\ =\ 0.5$ and $A\ =\ -0.3$ (with $c_{\,s}\ =\ \sqrt{1\ +\ A}$ for both peakons), respectively. This collision appeared to be a combination of the two previous collisions. After the interaction, the depression peakon sheds a small wavelet in front of the pulse, leading its propagation while the elevation peakon sheds an analogous wavelet behind, following the propagation of the main pulse. Figure~\ref{fig:mixedcol} shows the head-on collision between elevation and depression peakons.

Furthermore, we study the overtaking collision of two elevation and two depression peakons. Figure~\ref{fig:ovpeak1} shows the overtaking collision of two right-traveling elevation peakons of peak amplitudes $A\ =\ 2.5$ and $A\ =\ 1.1$ respectively. In this experiment, we take $\Delta x\ =\ 0.005$ and $\Delta t\ =\ 0.0005$ in the interval $[\,-100,\,100\,]\,$. The two peakons interact in an inelastic way and tails are generated during and after the interaction. It is noted that during the overtaking collisions of elevation peakons the peakons maintain some distance and exchange masses as in \textsc{Lax} category (\textit{a}). Figure~\ref{fig:ovpeak2} shows the overtaking collision of two right-traveling depression peakons of peak amplitudes $A\ =\ -0.1$ and $A\ =\ -0.5$ respectively. In this experiment, we take $\Delta x\ =\ 0.0025$ and $\Delta t\ =\ 0.00025\,$. Once again, counter-propagating tails have develop during the overtaking collision of these depression peakons while two peaks are observed for almost the whole interaction as in the \textsc{Lax} category (\textit{b}).

The last overtaking collision between an elevation and depression peakon is presented in Figure~\ref{fig:mixhead}. In this experiment, we use the same peakons as in the previous experiment but in this case both of them propagate to the right. The interaction is strong, especially for the depression peakon which evolve into a much smaller peakon while two more peakons are generated after the interaction. Also, other oscillatory structures appear. The initial elevation peakon evolve into a new elevation peakon of smaller amplitude.

It is noted that the tails generated during the interactions of traveling wave solutions for $B\ \approx\ 1/3$ all include a $N-$wave propagating in one direction. This indicates that as the effects of dispersion diminish, the dispersive tails behave similar to the non-dispersive tales of the critical case.

\begin{remark}
In order to explain the differences between the tails generated during the interactions discussed above, we study the properties of the \gSerre equations linearised around the trivial solution and in a reference frame $y\ =\ x\ -\ c_{\,s}\,t$ moving with the speed of a solitary wave, which is traveling to the right with speed $c_{\,s}\ >\ 0\,$. The equations then become
\begin{align}
  &(\partial_{\,t}\ -\ c_{\,s}\,\partial_{\,y})\,\eta\ +\ \partial_{\,y}\,u\ =\ 0\,, \label{eq:mref1}\\ 
  &(\partial_{\,t}\ -\ c_{\,s}\,\partial_{\,y})\,u\ +\ \partial_{\,y}\,\eta\ -\ \frac{1}{3}\;(\partial_{\,t}\ -\ c_{\,s}\,\partial_{\,y\,y})\,u\ -\ B\,\partial_{\,y\,y\,y}\,\eta\ =\ 0\,. \label{eq:mref2}
\end{align}
Plane wave solutions of the form $\eta\,(y,\,t)\ =\ a\,\ue^{\,{\rm i}\,(k\,y\ -\ \omega\,(k)\,t)}\,$, $u\,(y,\,t)\ =\ b\,\ue^{\,{\rm i}\,(k\,y\ -\ \omega\,(k)\,t)}$ of \eqref{eq:mref1} -- \eqref{eq:mref2} satisfy
\begin{align}
  & a\,(\omega\,(k)\ +\ c_{\,s}\,k)\,\ue^{\,{\rm i}\,(k\,y\ -\ \omega\,(k)\,t)}\ -\ b\,k\,\ue^{\,{\rm i}\,(k\,y\ -\ \omega\,(k)\,t)}\ =\ 0\,,  \label{eq:mref3}\\
  & a\,k\,(1\ +\ B\,k^{\,2})\,\ue^{\,{\rm i}\,(k\,y\ -\ \omega\,(k)\,t)}\ -\ b\,(\omega\,(k)\ +\ c_{\,s}\,k)\,\left(1\ +\ \frac{k^{\,2}}{3}\right)\ue^{\,{\rm i}\,(k\,y\ -\ \omega\,(k)\,t)}\ =\ 0\,. \label{eq:mref4}
\end{align}
The existence of nontrivial solution of \eqref{eq:mref3} -- \eqref{eq:mref4} implies that
\begin{equation*}
  (v\,(k)\ +\ c_{\,s})^{\,2}\ =\ \phi\,(k^{\,2})\,,
\end{equation*}
where $v\,(k)\ =\ \omega\,(k)/k$ and $\phi\,(x)\ =\ (1\ +\ B\,x)\,/\,(1\ +\ x/3)\,$, $x\ \geq\ 0\,$. It can be seen that $\phi\,(x)\ \leq\ \phi\,(0)$ when $B\ <\ 1/3$ and $\phi\,(x)\ \geq\ \phi\,(0)$ when $B\ >\ 1/3\,$. Therefore, we find 
\begin{align*}
  & -1\ <\ v\,(k)\ +\ c_{\,s}\ <\ 1\,,\qquad B\ <\ 1/3\,, \\
  & v\,(k)\ +\ c_{\,s}\ <\ -1\ \mbox{ or } v\,(k)\ +\ c_{\,s}\ >\ 1\,, \qquad B\ >\ 1/3\,.
\end{align*}
These relationships imply that in the case $B\ <\ 1/3\,$, where $c_{\,s}\ >\ 1\,$, the dispersive tails can propagate with speed $v\,(k)\ <\ 0$ and therefore follow the solitary waves, while in the case $B\ >\ 1/3\,$, where $c_{\,s}\ <\ 1\,$, the dispersive tails propagate with speed $v\,(k)\ >\ 0$ and therefore lead the solitary waves. In the critical case $B\ =\ 1/3$ there is no linear dispersion ($\phi\,(k^{\,2})\ =\ 1$) and therefore the tails are expected to propagate with speed $v\,(k)\ =\ -c_{\,s}\ \pm\ 1\,$. So if $c_{\,s}\ >\ 1$ then the generated tails should follow the traveling wave since $v\,(k)\ <\ 0$ but if $c_{\,s}\ <\ 1$ then tails can be generated in front and behind the main traveling wave. For example in Figure~\ref{fig:hocrit} where $c_{\,s}\ \approx\ 1.32\,$. Since $v\,(k)\ <\ 0$ the tails can propagate behind the solitary wave.  Analogous indications can be extracted for the case where $c_{\,s}\ <\ 1$ for depression peakons.
\end{remark}


\section{Conclusions}
\label{sec:concl}

Some effects of surface tension on gravity-capillary solitary waves of the \gSerre equations were presented. Head-on and overtaking collisions were studied for subcritical, critical, and supercritical values of the \textsc{Bond} number. The qualitative dynamical picture of the interactions for values of the \textsc{Bond} number $B\ <\ 1/3$ appeared to be very similar to the analogous interactions of solitary waves of the \textsc{Serre} equations that neglect surface tension. The maximum runup was smaller when surface tension was taken into account, and the effect of surface tension was stronger for larger amplitude solitary waves. On the other hand, the primary observed differences for solitary wave interactions with large values of \textsc{Bond} number $B\ >\ 1/3$ is the relative propagation of resultant dispersive tails, which propagate faster than the solitary waves as opposed to the slower when $B\ <\ 1/3\,$. We also studied the existence of solitary waves in the critical case $B\ =\ 1/3$ where the numerical experiments indicated the existence of solitary waves of elevation and depression that are reminiscent of peaked solitary waves of the \textsc{Camassa--Holm} equation. Analytical formulas for peakon solutions of the \gSerre equations were presented while their stability was explored. The numerical experiments related to the various nonlinear interactions of the traveling wave solutions of the \gSerre equations indicate that the model at hand is not integrable for any value of the \textsc{Bond} number $B\,$. Because the \gSerre equations are a fully nonlinear water wave model with weak or strong surface tension, the results obtained in this paper go beyond existing weakly nonlinear \textsc{Boussinesq} models and could be useful for understanding a variety of large-amplitude surface wave dynamics in thin fluid layers.


\subsection*{Acknowledgments}
\addcontentsline{toc}{subsection}{Acknowledgments}

M.~A.~\textsc{Hoefer} was partially supported by NSF CAREER DMS-1255422. D.~\textsc{Mitsotakis} was supported by the \textsc{Marsden} Fund administered by the Royal Society of \textsc{New Zealand}. The authors would also like to thank the anonymous referees for their valuable comments and suggestions that helped to improve the original manuscript.


\bigskip
\addcontentsline{toc}{section}{References}
\bibliographystyle{abbrv}
\bibliography{biblio}
\bigskip\bigskip

\end{document}